\documentclass[aps,prb,twocolumn,amsmath,amssymb,showpacs,superscriptaddress]{revtex4-2}

\usepackage{graphicx}
\usepackage{amssymb}
\usepackage{braket}
\usepackage{amsmath}
\usepackage{tabularx}
\usepackage{ltablex}
\usepackage{multirow}
\usepackage[dvipsnames]{xcolor}

\usepackage[colorlinks=true,
                linkcolor=blue,
	        citecolor=blue,
	        urlcolor=blue]{hyperref}

\newcommand{\trieste}{Dipartimento di Fisica, Universit\`a di Trieste, I-34151 Trieste, Italy}
\newcommand{\sissa}{Scuola Internazionale Superiore di Studi Avanzati (SISSA), I-34136 Trieste, Italy}

\usepackage{verbatim}

\AtBeginDocument{\usepackage{booktabs}}
\begin{document}

  \title{Theory of local \texorpdfstring{$\mathbb{Z}_2$}{} topological markers for finite and periodic two-dimensional systems}
  \date{\today}
  \author{Nicolas Baù}
  \affiliation{\trieste}
  \email{nicolas.bau@phd.units.it}
  \author{Antimo Marrazzo}
  \affiliation{\trieste}
  \affiliation{\sissa}
  \email{amarrazz@sissa.it}

  \begin{abstract}
    The topological phases of two-dimensional time-reversal symmetric insulators are classified by a $\mathbb Z_2$ topological invariant. Usually, the invariant is introduced and calculated by exploiting the way time-reversal symmetry acts in reciprocal space, hence implicitly assuming periodicity and homogeneity. Here, we introduce two space-resolved $\mathbb Z_2$ topological markers that are able to probe the local topology of the ground-state electronic structure also in the case of inhomogeneous and finite systems. The first approach leads to a generalized local spin-Chern marker, that usually remains well-defined also when the perpendicular component of the spin, $S_z$, is not conserved. The second marker is solely based on time-reversal symmetry, hence being more general. We validate our markers on the Kane-Mele model both in periodic and open boundary conditions, also in presence of disorder and including topological/trivial heterojunctions.
  \end{abstract}
  \maketitle
  \section{Introduction}\label{sec:intro}
    The topological phases of insulators are characterized by a bulk topological invariant, whose non-trivial character is manifested by gapless boundary modes. In the framework of two-dimensional systems, two main classes of topological insulators (TI) can be identified: The quantum anomalous Hall (also known as Chern) insulators (QAHI)~\cite{haldane_1988} and the time-reversal (TR) symmetric TIs, also known as quantum spin Hall insulators (QSHI)~\cite{kane_$z_2$_2005,kane_quantum_2005, bhzmodel_science_2006}. The topological phases of QAHIs are classified according to the Chern number $C\in\mathbb Z$, which signals the presence of chiral edge states protected by the non-trivial topology. As for QSHIs, TR symmetry is required and the symmetry-protected topological phase~\cite{kane_$z_2$_2005} is classified by an index $\nu\in\mathbb Z_2$. QSHIs display an odd number of helical edge states and edge currents that are spin-momentum locked.

    Within periodic boundary conditions (PBC), the presence of a non-zero topological invariant is related to the impossibility of constructing a smooth gauge in the whole Brillouin zone (BZ)~\cite{topological_obstruction_chern, soluyanov_wannier_z2_2011} and represents a topological obstruction. Topological invariants are usually defined as global quantities of the system, and in general they cannot be evaluated as the expectation value of some operator. The task further complicates when considering symmetry-protected topological phases, as it becomes necessary to include the protecting symmetries in the definition of the topological invariant. Examples for TR-symmetric insulators include the concept of TR polarization~\cite{ku_kane_trpolarization}, the parity of TR invariant momenta~\cite{fu_kane_inv}, or the flow of hybrid Wannier functions centers~\cite{soluyanov_wannier_z2_2011, yu_z2_wf_prb_2011, z2pack}. For electronic structure simulations of solids PBCs are usually chosen, and symmetries are typically directly incorporated in reciprocal space~\cite{fu_kane_inv, ku_kane_trpolarization}, rather than real space.
    However, by construction, these global formulations target homogeneous and periodic systems. In the case of periodic and disordered systems a number of strategies have been developed to compute the global topological invariant, such as the single-point sampling of the BZ in the large supercell framework~\cite{ceresoli_sp_prb_2007,favata_es_2023}, the Bott index~\cite{Loring_2011, huang_prb_2018}, the structural spillage~\cite{grushin_prr_2023}, and methods based on the scattering matrix~\cite{PhysRevB.85.165409, PhysRevB.83.155429} or on the non-commutative index theorem~\cite{Avron1994, AVRON1994220, 10.1063/1.5026964}.

    When dealing with finite or inhomogeneous systems, global invariants are of no avail, and one must adopt a space-resolved approach, as done for instance in Refs.~\cite{bianco_prb_2011, kpm_lcm_2020, confinedhoftadter_gebert_2020, disorderchern_prl_2020, dornellas_prb_2022, mildner_2023}. The ability to sample these quantities locally in real space can be particularly useful when dealing, for instance, with amorphous systems~\cite{grushin_amorphousti_2020}. The long-range order\textemdash that usually allows calculating the topological invariant via reciprocal space formulations\textemdash is lost, and only on the local scale the system appears to retain an ordered arrangement.
    Since the topological invariant is related to the organization of electrons in the ground state, in principle it does not require the existence of long-range order or translational invariance. As the ground-state electronic structure is ``nearsighted''~\cite{kohn_prl_1996}, it should be possible to probe the local topology of a region just with the knowledge of its neighborhood in real space. A simple and relevant scenario is a heterostructure composed of two regions with different topology, such as an insulator in a topological phase at the interface with a trivial insulator (or vacuum). There, the meaning of a global invariant is ambiguous, since a single invariant cannot capture the individual bulk topology of the two regions and the presence of metallic edge states at the interface between them.

    In this work, we develop space-resolved formulas for the $\mathbb Z_2$ topological invariant that can be used to probe the local topology of inhomogeneous and non-crystalline two-dimensional systems. Over the past years, a number of local markers have been developed to investigate the presence of topological phases in different systems. Examples include the local Chern marker~\cite{bianco_prb_2011}, that has been used to study also QSHIs~\cite{assuncao_lcmBHZ_2024}, the mirror-Chern marker~\cite{kpm_lcm_2020}, a spin-Chern marker based on the spin-Berry curvature~\cite{weichen_spincurvature_2023}, a definition of the local $\mathbb Z_2$ invariant based on a flux-insertion-induced spectral flow~\cite{li_prb_2019}, a local marker to identify crystalline topological phases~\cite{lcti_prl_2024} and a layer-resolved spin-Chern marker for 3D TIs~\cite{layerlscm_natcomm_2024}. Recently, a marker to investigate higher-order topological phases has also been proposed~\cite{higherordermarker_npj_2023}. In particular, the archetypal case of a local Chern marker for two-dimensional QAHIs has been investigated in Ref.~\cite{bianco_prb_2011} within open boundary conditions (OBCs) and in Ref.~\cite{pbclcm} within PBCs. The local Chern marker~\cite{bianco_prb_2011} can also be accessed experimentally by measuring local circular dichroism~\cite{Pozo_PRL_2019, Marsal_PNAS_2020, Molignini_scipost_2023}. Moreover, a generalization of the concept of local marker in odd dimensions $D$ has been introduced in Ref.~\cite{hannukainen_2022} for both $\mathbb Z$-phases (local Chiral marker) and $\mathbb Z_2$-phases (local Chern-Simons marker). That is done by introducing a $(D+1)$-dimensional local Chern marker, extending the one by Bianco-Resta~\cite{bianco_prb_2011} to higher dimensions, together with a family of parametrized projectors (where the parameter acts as the additional dimension) that have to be integrated in order to get a local marker when $D$ is odd. In particular, the invariant is $\mathbb Z$-valued if the single-particle density matrix obeys a chiral constraint, and $\mathbb Z_2$ otherwise~\cite{tenfoldway_2010}. Recently, a universal topological marker, valid in any dimension and symmetry class, has been proposed in Ref.~\cite{wei_prb_2022}. This marker has been successfully tested on various systems belonging to different symmetry classes, but in the most general case (of interest for this work) of two-dimensional class AII systems with Rashba spin-orbit coupling (SOC), such as the Kane-Mele model~\cite{kane_quantum_2005}, a general and practical formulation of the local $\mathbb Z_2$ marker that can always be applied is still missing. Indeed, Rashba SOC (that can be intrinsic or induced, for instance, by an electric field perpendicular to the system) violates $S_z$ conservation, so one can no longer consider spins to be ``up'' or ``down'', and must resort to a different strategy to calculate the $\mathbb Z_2$ invariant.
    
    There are many available methods to compute the global topological invariant of QSHIs by making explicit use of TR symmetry~\cite{fu_kane_inv, ku_kane_trpolarization}. For our purposes, two particularly noteworthy approaches are the spin-Chern number~\cite{prodan_2009} and the calculation of hybrid Wannier charge centers~\cite{soluyanov_prb_2011, z2pack}. The underlying common strategy between these two methods is to split the Hilbert space spanned by the occupied states as a sum of subspaces such that the individual Chern number associated to each subspace is an integer~\cite{soluyanov_prb_2011,z2pack}. That can be done in an arbitrary way, but the individual Chern numbers are meaningful only when such splitting is realized according to the symmetries protecting the topological phase. In QSHIs, for instance, one can use the projected spin operator~\cite{prodan_2009} or the TR symmetry~\cite{soluyanov_2012}: The two subspaces obtained individually break TR symmetry and can host a non-zero Chern number. Hence, calculating the $\mathbb Z_2$ topological invariant translates to the problem of identifying a suitable partitioning of the occupied Hilbert space where it is possible to compute integer individual Chern numbers. However, performing such partitioning might be rather non-trivial in practice. Indeed, it has been discussed~\cite{bellissard_noncommutative,prodan_2009,prodan_2010} that an exponentially localized projector commuting with lattice translations is a sufficient condition to yield integer individual Chern numbers that are invariant with respect to small perturbations of the system.
    
    Here, we propose two formulations for local $\mathbb Z_2$ invariants: The first leverages the spin-Chern number introduced by Prodan~\cite{prodan_2009}, while the second is based solely on TR symmetry, in the spirit of the work by Soluyanov and Vanderbilt~\cite{soluyanov_phd, soluyanov_2012}. These two strategies are used to define markers for finite samples in OBCs as well as for large supercells in PBCs (where the BZ can be effectively sampled by a single point, typically $\Gamma$). Our approach is particularly useful when dealing with non-homogeneous systems, such as disordered samples and heterostructures, giving the possibility of inspecting the topology locally in real space.

    The paper is organized as follows. In Sec.~\ref{sec:obc_pbc_lcm}, we recap on some relevant results from the literature, namely the OBC and PBC local Chern markers developed for QAHIs. In Sec.~\ref{sec:z2obc}, we first discuss the local spin-Chern marker and then introduce the local $\mathbb Z_2$ marker based on TR symmetry within OBCs, that we further generalize to PBCs in Sec.~\ref{sec:z2pbc}. Then, in Sec.~\ref{sec:methods} we discuss the tight-binding models used to benchmark our markers and introduce a generalized smearing, which is particularly useful when dealing with heterostructures. In Sec.~\ref{sec:results}, we present numerical results and compare the different methods introduced. Finally, we summarize our results and conclusions in Sec.~\ref{sec:conclusions}.

  \section{Local Chern markers}\label{sec:obc_pbc_lcm}
    In this section we review the OBC and PBC formulations of the LCM introduced in Refs.~\cite{bianco_prb_2011, pbclcm}. Let us consider a QAHI with $N_{occ}$ occupied bands, and let $\ket{u_{n\mathbf k}}=e^{-i\mathbf k\cdot\mathbf r}\ket{\psi_{n\mathbf k}}$ be the periodic part of Bloch functions. The Hamiltonian of a QAHI breaks TR symmetry, and its topological invariant is the Chern number:
    \begin{eqnarray}\label{eq:chern_number}
      C=-\frac{1}{\pi}\mathrm{Im}\sum_{n=1}^{N_{occ}}\int_{BZ}d\mathbf k \braket{\partial_{k_x}u_{n\mathbf k}|\partial_{k_y} u_{n\mathbf k}}.
    \end{eqnarray}
    Since the Chern number is defined in the primitive cell, it represents a global quantity meaningful only for pristine and homogeneous systems within PBCs. Bianco and Resta~\cite{bianco_prb_2011} showed that, starting from Eq.~\ref{eq:chern_number}, it is possible to derive a local Chern marker (LCM):
    \begin{align}\label{eq:obclcm_1}
      \mathfrak C(\mathbf r)&=-4\pi\mathrm{Im}\braket{\mathbf r | \mathcal P x (\mathbb I-\mathcal P)y|\mathbf r}\\
      &=4\pi\mathrm{Im}\braket{\mathbf r|\mathcal P[x,\mathcal P][y, \mathcal P]|\mathbf r}\label{eq:obclcm_2}
    \end{align}
    such that its macroscopic average over a small region in real space describes the local topology. In a pristine system, that reduces to the trace per unit area $\mathrm{Tr_A}$ in the region of interest, and results in the local Chern number of the system. In Equations~\ref{eq:obclcm_1} and~\ref{eq:obclcm_2}, $x$ and $y$ are the Cartesian components of the position operator $\mathbf r$, and $\mathcal P=\sum_{n=1}^{N_{occ}}\ket{u_n}\bra{u_n}$ is the ground-state projector. The key feature of this formulation is that, by being expressed as a trace, the Chern number can be evaluated locally in real space, that makes it particularly suited to study disordered and inhomogeneous systems. The LCM defined in Eq.~\ref{eq:obclcm_2} is also related to the geometrical intrinsic contribution to the local anomalous Hall conductivity in metals~\cite{marrazzo_prb_2017, rauch_ahc_pbc2018}. So defined, the LCM of Eq.~\ref{eq:obclcm_2} could, in principle, be applied in both PBCs and OBCs, since the operators $[r_{\alpha}, \mathcal P]$ are well-defined in both cases, despite $\mathbf r$ being an illegitimate operator within PBCs~\cite{resta_prl_1998}. However, when dealing with numerical simulations, the operators $[r_{\alpha}, \mathcal P]$ are not directly accessible and the usual way to construct them is to multiply the ``standard'' position operator $\mathbf r$ with $\mathcal P$, thus limiting this method to the OBC case only~\cite{pbclcm}. Equation~\ref{eq:obclcm_2} shows very clearly that a finite system is always globally trivial in a certain sense, as its Chern number is always zero~\cite{bianco_prb_2011} being the trace over the whole sample of commutators of matrices with finite dimensions.
    
   We can recover the local topology also within PBCs \cite{pbclcm}, by leveraging the single-point formulation of the Chern number in the large supercell limit~\cite{ceresoli_sp_prb_2007}:
    \begin{eqnarray}\label{eq:chern_singlepoint}
      C^{(asym)}=-\frac{1}{\pi}\mathrm{Im}\sum_{n=1}^{N_{occ}}\braket{\tilde u_{n\mathbf b_1}|\tilde u_{n\mathbf b_2}},
    \end{eqnarray}
    where $\mathbf b_{1,2}$ are the reciprocal lattice vectors and $\ket{\tilde u_{n\mathbf b_j}}$ are the ``dual'' vectors of $\ket{u_{n\Gamma}}$, the latter are the periodic part of the Hamiltonian eigenstates at $\Gamma$. The dual states are defined as:
    \begin{eqnarray}
      \ket{\tilde u_{n\mathbf b_j}}=\sum_{m=1}^{N_{occ}}S^{-1}_{mn}(\mathbf b_j)e^{-i\mathbf b_j\cdot\mathbf r}\ket{u_{m\Gamma}},
    \end{eqnarray}
    where $S_{mn}(\mathbf b_j)=\braket{u_{m\Gamma}|e^{-i\mathbf b_j\cdot\mathbf r}|u_{n\Gamma}}$ is the overlap matrix between the states at $\Gamma$ and the ones at $\mathbf b_j$ once the periodic gauge is imposed. The dual vectors satisfy $\braket{\tilde u_{n\mathbf b_j}|u_{m\Gamma}}=\delta_{nm}$ and represent, in the limit of a large supercell, the vectors obtained from a parallel-transport procedure from $\Gamma$ to $\mathbf b_j$~\cite{souza_covariantderivative_prb_2004,favata_es_2023}. The single-point formulation is based on a discretization of the covariant derivative, and depending on whether this is approximated by forward or symmetric finite differences, the asymmetric formula [Eq.~\ref{eq:chern_singlepoint}] or the symmetric one [Eq.~\ref{eq:chern_singlepoint_sym}] can be obtained, respectively:
    \begin{align}
      C^{(sym)}= -\frac{1}{4\pi}\mathrm{Im}\sum_{n=1}^{N_{occ}}\big( &\bra{\tilde u_{n\mathbf b_1}}-\bra{\tilde u_{n-\mathbf b_1}} \big)\cdot\nonumber\\ &\cdot\big( \ket{\tilde u_{n\mathbf b_2}}-\ket{\tilde u_{n-\mathbf b_2}} \big).\label{eq:chern_singlepoint_sym}
    \end{align}
    As shown in Ref.~\cite{pbclcm}, from Eqs.~\ref{eq:chern_singlepoint} and~\ref{eq:chern_singlepoint_sym} the corresponding PBC LCMs can be defined as:
    \begin{align}\label{eq:pbclcm}
      &\mathcal C^{(asym)}(\mathbf r)=-\frac{1}{2\pi}\mathrm{Im}\braket{\mathbf r|[\mathcal P_{\mathbf b_1},\mathcal P_{\mathbf b_2}]\mathcal P_{\Gamma}|\mathbf r}
    \end{align}
    and
    \begin{align}
      &\mathcal C^{(sym)}(\mathbf r)=-\frac{1}{8\pi}\mathrm{Im}\bra{\mathbf r}\Big([\mathcal P_{\mathbf b_1},\mathcal P_{\mathbf b_2}]+[\mathcal P_{-\mathbf b_1},\mathcal P_{-\mathbf b_2}]-\nonumber \\
      &\hspace{1cm}-[\mathcal P_{-\mathbf b_1},\mathcal P_{\mathbf b_2}]-[\mathcal P_{\mathbf b_1},\mathcal P_{-\mathbf b_2}]\Big)\mathcal P_{\Gamma}\ket{\mathbf r}\label{eq:pbclcm_sym}
    \end{align}
    where $\mathcal P_{\mathbf b_j}=\sum_{n=1}^{N_{occ}}\ket{\tilde u_{n\mathbf b_j}}\bra{\tilde u_{n\mathbf b_j}}$ and $\mathcal P_{\Gamma}=\sum_{n=1}^{N_{occ}}\ket{u_{n\Gamma}}\bra{u_{n\Gamma}}$ are ground-state projectors. In particular, Eq.~\ref{eq:pbclcm} derives from the asymmetric single-point formulation [Eq.~\ref{eq:chern_singlepoint}], while Eq.~\ref{eq:pbclcm_sym} derives from the symmetric one [Eq.~\ref{eq:chern_singlepoint_sym}]. As discussed for the OBC case, also in PBCs the local topology is described by the macroscopic average of the LCM over a small region in real space. A key feature of this formulation is that the position operator appears only through the exponential $e^{-i\mathbf b_j\cdot \mathbf r}$, thus resulting in a legitimate operator also within PBCs.
    The PBC LCM offers a simple picture of the topological obstruction in non-crystalline systems: In the trivial ($C=0$) case, one can always have $[\mathcal P_{\mathbf b_1},\mathcal P_{\mathbf b_2}]=0$ due to the existence of a smooth gauge in the whole BZ. In the topological case, however, this construction is impossible, and a non-zero (local) Chern number arises from the fact that $\mathcal P_{\mathbf b_1}$ and $\mathcal P_{\mathbf b_2}$ (locally) do not commute, also in inhomogeneous systems.
    If one considers the case of a finite sample inside a much larger supercell with PBCs, the LCM of Eq.~\ref{eq:pbclcm} can be expanded in powers of the linear dimension of the supercell $L$: In the limit of $L\rightarrow\infty$, the expansion converges to the LCM of Eq.~\ref{eq:obclcm_2}. In addition, while the OBC LCM [Eq.~\ref{eq:obclcm_2}] vanishes upon tracing over the entire system, the PBC LCM [Eqs.~\ref{eq:pbclcm},\ref{eq:pbclcm_sym}] results in the single-point Chern number~\cite{ceresoli_sp_prb_2007}, even in the disordered case.
    Both the OBC and PBC formulations of the LCM can be applied to study the local topology of disordered and amorphous materials, as well as inhomogeneous systems such as trivial/topological heterojunctions and superlattices.

  \section{Local \texorpdfstring{$\mathbb{Z}_2$}{} markers for finite systems}\label{sec:z2obc}
    As mentioned in Sec.~\ref{sec:intro}, there are many equivalent methods available to compute the global topological $\mathbb Z_2$ invariant of a QSHI. However, in general, it is not a trivial task to remove the notion of reciprocal space in their definition, as they often rely on some symmetries of the BZ under the TR operator $\Theta$. Moreover, $\Theta$ is antiunitary, so unlike other unitary symmetries protecting topological phases (like the mirror operator~\cite{mirrorcherndef_prb_2008}), it cannot be diagonalized to get two Chern subspaces with well-defined symmetry labels, and other strategies have to be developed.

    Because of Kramers' theorem, each eigenvalue of a TR symmetric Hamiltonian is at least doubly degenerate. Hence, only half of the occupied states is really needed to compute the topological invariant, since the other half can be obtained by symmetry. One can then split the occupied Hilbert space into two Chern subspaces and obtain two set of states (built in general as linear combination of the occupied eigenstates of the Hamiltonian) that are mapped onto each other by TR symmetry. For instance, one can split the Kramers-degenerate eigenstates of the Hamiltonian such that, for each vector assigned to a subspace, its TR partner is assigned to the other. By doing so, in each subspace TR symmetry is broken by construction, so it is possible and meaningful to compute its individual Chern number~\cite{soluyanov_prb_2011,z2pack}. Moreover, since TR symmetry forces $C=0$, the two subspaces must be characterized by opposite individual Chern numbers.
    Here we promote the individual Chern number to a local individual Chern marker, and use Eq.~\ref{eq:obclcm_2} on the states belonging to just one of the two subspaces. Obtaining a decomposition of the occupied Hilbert space that leads to an integer individual Chern number, however, is not a trivial task. In fact, for the individual Chern numbers to be well-defined, we need exponentially localized projectors onto their corresponding subspaces ~\cite{bellissard_noncommutative, prodan_2009, prodan_2010}. Here, we propose two different methods to realize the splitting of the occupied states into two Chern subspaces characterized by exponentially localized projectors, first via the projected spin operator $\mathcal P S_z\mathcal P$ and then by using the TR symmetry operator $\Theta$, leading to two distinct local markers of the $\mathbb Z_2$ topology.

    \subsection{Local OBC spin-Chern marker}\label{sec:z2obc_a}
      We start by considering a Hamiltonian $\mathcal H$ for which $[\mathcal H, S_z]=0$, where $S_z$ is the spin operator along the $z$-direction. Since $\mathcal H$ commutes with $S_z$, it can be decoupled into spin sectors that are mapped onto each other by TR symmetry. Such splitting allows the definition of two individual Chern numbers $C_{\uparrow,\downarrow}$, physically related to the quantized spin Hall conductivity~\cite{rauch_prb_2020}, from which the $\mathbb Z_2$ invariant of the system can be computed as:
      \begin{eqnarray}\label{eq:global_scn}
        \nu=\frac{C_{\uparrow}-C_{\downarrow}}{2}\mod 2,
      \end{eqnarray}
      where the two individual Chern numbers can be obtained by integrating the Berry curvature of the relative spin-up or spin-down subspace. The states composing the two spin sectors can be selected by computing the ground-state projector $\mathcal P(\mathbf k)$ and diagonalizing $\mathcal P(\mathbf k)S_z\mathcal P(\mathbf k)$, whose spectrum has eigenvalues $\pm\frac{1}{2}$. Remarkably, Prodan~\cite{prodan_2009} proved that this strategy holds even if $[\mathcal H, S_z]\neq 0$ as long as the operator $\mathcal P(\mathbf k)S_z\mathcal P(\mathbf k)$ displays a gap in its spectrum. In this case, if one diagonalize $\mathcal P(\mathbf k)S_z\mathcal P(\mathbf k)$, its eigenvalues $s_{\lambda}$ will spread symmetrically around zero in the interval $[-\frac{1}{2}, \frac{1}{2}]$, and the projectors onto the positive ($\sigma=+$) and negative ($\sigma=-$) eigenvalues can be defined:
      \begin{eqnarray}
        \mathcal P_{\sigma}(\mathbf k)=\sum_{\lambda:\mathrm{sign}(s_{\lambda})=\sigma}\ket{\phi_{\lambda\mathbf k}}\bra{\phi_{\lambda\mathbf k}},
      \end{eqnarray}
      where $\ket{\phi_{\lambda\mathbf k}}$ are the eigenstates of $\mathcal P(\mathbf k)S_z\mathcal P(\mathbf k)$ with eigenvalue $s_{\lambda}$. The existence of a spectral gap in the projected spin operator allows retaining exponentially localized projectors $\mathcal P_{\sigma}(\mathbf k)$, so that an integer individual Chern number can still be defined. Moreover, the splitting of the eigenstates according to the projected spin operator satisfy:
      \begin{eqnarray}
        \mathcal P(\mathbf k)=\mathcal P_{\sigma}(\mathbf k)+\mathcal P_{-\sigma}(\mathbf k)=\mathcal P_{\sigma}(\mathbf k)+\Theta \mathcal P_{\sigma}(-\mathbf k)\Theta^{-1}
      \end{eqnarray}
      so, due to TR symmetry, it must hold that $C_+=-C_-$, and the $\mathbb Z_2$ invariant can be computed as a spin-Chern number, as in Eq.~\ref{eq:global_scn}, by substituting $C_{\uparrow}\rightarrow C_+$ and $C_{\downarrow}\rightarrow C_-$. It should be noted that other definitions of the spin-Chern number exists, for instance as the integral of the spin-Berry curvature~\cite{weichen_spincurvature_2023}. However, while Prodan's spin-Chern number~\cite{prodan_2009} is quantized, the integral of the spin-Berry curvature is only an approximate indicator of the topology of the system, although experimentally relevant since related to the spin-resolved local circular dichroism~\cite{weichen_spincurvature_2023}.

      Within OBCs the BZ does not exist: A local spin-Chern marker (LSCM) can then be obtained by applying the LCM formula [Eq.~\ref{eq:obclcm_2}] to each of the individual Chern numbers appearing in Eq.~\ref{eq:global_scn}. By doing so, two individual local Chern markers can be computed:
      \begin{eqnarray}
        \mathfrak C_{\sigma}(\mathbf r)=4\pi\mathrm{Im}\braket{\mathbf r|\mathcal P_{\sigma}[x,\mathcal P_{\sigma}][y,\mathcal P_{\sigma}]|\mathbf r}
      \end{eqnarray}
      which, substituted into Eq.~\ref{eq:global_scn}, results in a LSCM defined as
      \begin{eqnarray}\label{eq:lscm}
        \nu(\mathbf r)=\frac{\mathfrak C_+(\mathbf r)-\mathfrak C_-(\mathbf r)}{2}\mod 2,
      \end{eqnarray}
      whose macroscopic average is a local $\mathbb Z_2$ invariant. Being expressed as a trace, the LSCM can be evaluated in real space as for the OBC LCM. A similar strategy can also be used to define a layer-resolved spin-Chern marker in 3D systems~\cite{layerlscm_natcomm_2024}.
      We stress again that such formulation is useful mostly within OBCs, as the general way to calculate the operators $[r_{\alpha}, \mathcal P_{\sigma}]$ requires the explicit evaluation of the terms $r_{\alpha}\mathcal P_{\sigma}$ and $\mathcal P_{\sigma}r_{\alpha}$, which are ill-defined within PBCs. We will discuss how to generalize these local markers to the PBC framework in Sec.~\ref{sec:z2pbc}. As stated before, the LSCM remains well-defined whenever the projected spin operator displays a spectral gap, possibly also for systems breaking TR symmetry. However, in general, the existence of a gap in $\mathcal PS_z\mathcal P$ is not guaranteed, as it is not a fundamental property of QSHIs. Hence, the LSCM is a valid approach but not, in principle, a truly general formulation of the $\mathbb Z_2$ invariant.

    \subsection{Local OBC \texorpdfstring{$\mathbb{Z}_2$}{} marker based on time-reversal symmetry}\label{sec:z2obc_b}
      QSHIs are characterized by the TR symmetry $\Theta$, which forces the Chern number of the system to vanish, and the classification of topological phases to be $\mathbb Z_2$, resulting in a symmetry-protected topological phase. Hence, it seems reasonable that a general definition of the $\mathbb Z_2$ topological invariant should exploit TR symmetry explicitly, similarly to what has been done, for instance, for the mirror-Chern number~\cite{mirrorcherndef_prb_2008, mirrorchern_prb_2021}. However, the TR operator $\Theta$ is antiunitary and it cannot be diagonalized to define a topological invariant. Nonetheless, we can still split the occupied states into two Chern subspaces using $\Theta$ thanks to Kramers' theorem. A similar strategy was proven to be successful in computing the global invariant of a homogeneous and periodic QSHI with only two occupied bands~\cite{soluyanov_2012}. The calculation of the topological invariant, in this case, is achieved by enforcing the parallel-transport gauge in almost all the BZ: If one insists in having a periodic and TR-symmetric gauge, this cannot be smooth due to the topological obstruction arising in the topological phase. To evaluate the $\mathbb Z_2$ invariant one can impose a TR symmetric gauge that is smooth on the BZ cylinder (but not on the BZ torus, i.e., a non-periodic gauge), disentangling the occupied bands into two Chern manifolds, each characterized by its own integer individual Chern number~\cite{soluyanov_2012}.

      Within OBCs, the parallel transport procedure becomes irrelevant, and the only meaningful step is the TR constraint. Numerical diagonalizations carry random global phases for each eigenvector, so to impose a TR symmetric gauge we need to select an initial state in the degenerate subspace and explicitly evaluate its TR partner. If the dimension of the degenerate subspace is greater than two, then an orthonormalization procedure is needed. This procedure can be repeated until the number of vectors is equal to the dimension of the degenerate subspace. By doing so, the eigenstates of the Hamiltonian are split in two orthogonal subsets (identified by the projectors $\mathcal P_1$ and $\mathcal P_2$) that individually break TR, and that can be mapped onto each other by symmetry. That allows us to evaluate the individual local Chern numbers of the two subsets, and obtain a local $\mathbb Z_2$ marker. However, this construction in general does not result in exponentially localized projectors $\mathcal P_{1,2}$, and their decay properties in the bulk depend on the choice of the eigenstates assigned to each subspace.

      One way to obtain exponentially localized projectors is to build the maximally localized Wannier functions (MLWF) of the system~\cite{marzari_prb_1997, marzari_wf_review} (also known as Boys orbitals in OBCs~\cite{boys_obcwf}). TR symmetric systems are characterized by a vanishing Chern number, which in principle should allow computing exponentially localized Wannier functions (WF). In practice, however, the simplest method to compute them is via a projection procedure~\cite{marzari_wf_review}, which is ensured to fail in the topological phase if the trial states are TR symmetric~\cite{wannier_z2_dimensions_soluyanov}. This is a consequence of the topological obstruction, that forbids the existence of a smooth and TR-symmetric gauge in the topological phase~\cite{soluyanov_2012}.
      Within OBCs, a set of trial functions $g_n(\mathbf r)$ can be chosen based on chemical intuition. Then, these are used to project the $J$ occupied orbitals:
      \begin{eqnarray}\label{eq:wfprojection}
        \ket{\phi_n}=\sum_{m=1}^{J}\ket{\psi_m}\braket{\psi_m|g_n}.
      \end{eqnarray}
      The trial orbitals $\ket{\phi_n}$ are then rotated by using the overlap matrix $S_{mn}=\braket{\phi_m|\phi_n}$, obtaining ``projected'' WFs:
      \begin{eqnarray}\label{eq:wf_projection}
        \ket{\tilde w_n}=\sum_{m=1}^J\ket{\phi_m}S^{-1}_{mn}.
      \end{eqnarray}
      If MLWFs are sought, one needs to find the unitary rotation $U$ such that $\ket{w_m}=\sum_nU_{nm}\ket{\tilde w_n}$ are the WFs that minimize the quadratic spread
      \begin{eqnarray}\label{eq:spread}
        \Omega=\sum_n\left[\braket{w_n|\mathbf r^2|w_n}-\left|\braket{w_n|\mathbf r|w_n}\right|^2\right].
      \end{eqnarray}
      MLWFs can then be split in two sets by using TR symmetry as explained above. This procedure allows us to define a local $\mathbb Z_2$ marker (LZ2M) that is uniquely based on TR symmetry as
      \begin{eqnarray}
        \Delta(\mathbf r)=\frac{\mathfrak C_1(\mathbf r)-\mathfrak{C}_2(\mathbf r)}{2}\mod 2,
      \end{eqnarray}
      where $\mathfrak C_1(\mathbf r)$ and $\mathfrak C_2(\mathbf r)$ are computed via Eq.~\ref{eq:obclcm_2} with the projectors $\mathcal P_1$ and $\mathcal P_2$ respectively. It is worth noting that, as the LZ2M is based entirely on TR symmetry, its formulation does not rely on the existence of a spectral gap in any operator (but the Hamiltonian), thus being more general with respect to the LSCM. However, the validity of the LZ2M strongly depends on the choice of the trial functions $\ket{g_n}$ introduced in Eq.~\ref{eq:wfprojection}. Specifically, as explained in Sec.~\ref{sec:trial_projections}, in order to avoid the topological obstruction, the trial functions $\ket{g_n}$ should not be related by TR symmetry. This condition, in turn, implies that $\mathcal P\neq\mathcal P_1+\Theta\mathcal P_1\Theta^{-1}$. However, since we want to split the occupied Hilbert space into two subsets that represents the whole space and are related by TR symmetry (to compute the $\mathbb Z_2$ invariant from Chern numbers), we need the two projectors $\mathcal P$ and $\mathcal P_{\Theta}=\mathcal P_1+\Theta\mathcal P_1\Theta^{-1}$ to be as close as possible~\cite{z2pack}. For this reason, we require that the choice of the trial projection, discussed in Sec.~\ref{sec:trial_projections}, should also minimize the spillage between the projectors $\mathcal P$ and $\mathcal P_{\Theta}$, defined as~\footnote{\label{footnote:1} The spillage can be interpreted as a measure of how much two projectors describe the same subspace, or equivalently, a distance between manifolds. Since the trace of $\mathcal P$ is equal to the number of occupied states $N_{occ}$, this formulation of the spillage is equivalent to the more common expression in terms of the projector onto empty states: $\mathrm{Tr}[(\mathcal P-\mathcal P_{\Theta})^2]/(2N_{occ})=\mathrm{Tr}[\mathcal P\mathcal Q_{\Theta}]/N_{occ}=\mathrm{Tr}[\mathcal Q\mathcal P_{\Theta}]/N_{occ}$. From these forms it is easier to see that, for $\gamma=0$, the projectors describe the same subspace, while if $\gamma=1$, the two projectors are orthogonal and thus describe orthogonal subspaces.}:
      \begin{eqnarray}\label{eq:spillage_projectors}
        \gamma = \frac{1}{2N_{occ}}\mathrm{Tr}\big[ (\mathcal P - \mathcal P_{\Theta})^2 \big].
      \end{eqnarray}
      That allows approximating the manifold of occupied states as the sum of two ``Chern subspaces'' in which we can evaluate individual Chern numbers. Note that we can exchange the states from one subspace to the other without changing the $\mathbb Z_2$ topological invariant. In fact, as long as the states are mapped onto each other by TR symmetry, they carry opposite contributions and the $\mathbb Z_2$ invariant does not change~\cite{z2pack}.

      We also observe that the steepest descent procedure to compute MLWFs does not particularly improve the value of the topological marker, and projections-only WFs are sufficient to obtain an accurate map of the local topology. To be more precise, in this case one should not refer to the states obtained after the projection procedure as WFs, since the minimization of the spillage (which translates to a requirement of minimal TR symmetry breaking in the gauge) does not allow obtaining truly exponentially localized WFs, as it will be shown through numerical simulations in Sec.~\ref{sec:convergence}. These orbitals are somehow localized in real space but also characterized by a seemingly divergent spread in the thermodynamic limit: In the following we will refer to them as \textit{quasi}-WFs (qWF). Notably, a certain degree of localization in real space\textemdash not necessarily asympotically exponential\textemdash appears to be a sufficient condition to define bona fide local markers. Indeed, qWFs yield projectors $\mathcal P_{1,2}$ that might be exponentially localized only at short distances in the topological phase. Still, short-range exponential localization seems to be a sufficient condition to calculate integer local individual Chern numbers $C_{1,2}$ through Eq.~\ref{eq:obclcm_2}.

  \section{Local \texorpdfstring{$\mathbb{Z}_2$}{} markers for periodic systems}\label{sec:z2pbc}
      \subsection{Local PBC spin-Chern marker}
      Within PBCs, following the same strategy developed in Sec.~\ref{sec:z2obc_a}, the individual Chern numbers $C_{\pm}$ introduced in Eq.~\ref{eq:global_scn} can be promoted to local individual Chern markers using the PBC LCM [Eq.~\ref{eq:pbclcm}]. A complementary, and equivalent, way to look at this strategy is that we derive from single-point spin-Chern numbers introduced in Ref.~\cite{favata_es_2023} the corresponding local markers. In the limit of a very large supercell, the BZ shrinks to the $\Gamma$-point only, so just one diagonalization of the Hamiltonian is required. Defining $\mathcal P_{\Gamma}=\sum_{n=1}^{N_{occ}}\ket{u_{n\Gamma}}\bra{u_{n\Gamma}}$ the ground-state projector, we can diagonalize the projected spin operator $\left(\mathcal P_{\Gamma}S_z\mathcal P_{\Gamma}\right)\ket{\phi_{\lambda\Gamma}}=s_{\lambda}\ket{\phi_{\lambda\Gamma}}$, and construct the dual states for each spin sector:
      \begin{eqnarray}
        \ket{\tilde{\phi}_{\lambda\mathbf b_j}}=\sum_{\mu:\mathrm{sign}(s_{\mu})=\sigma}S_{\mu\lambda}^{-1}(\mathbf b_j)e^{-i\mathbf b_j\cdot\mathbf r}\ket{\phi_{\mu\Gamma}}
      \end{eqnarray}
      where $\sigma=\pm$ identifies the spin sector with overlap matrix $S_{\mu\lambda}(\mathbf b_j)=\braket{\phi_{\mu\Gamma}|e^{-i\mathbf b_j\cdot\mathbf r}|\phi_{\lambda\Gamma}}$. Then, we define the projectors
      \begin{eqnarray}
        \mathcal P_{\mathbf b_j}^{\pm} = \sum_{\lambda:\mathrm{sign}(s_{\lambda})=\pm}\ket{\tilde{\phi}_{\lambda\mathbf b_j}}\bra{\tilde{\phi}_{\lambda\mathbf b_j}},
      \end{eqnarray}
      that can be used to introduce two individual LCMs for positive and negative eigenvalues of $\mathcal P_{\Gamma}S_z\mathcal P_{\Gamma}$:
      \begin{align}\label{eq:ilcm_spin}
        &\mathcal C_{\pm}^{(asym)}(\mathbf r)=-\frac{1}{2\pi}\mathrm{Im}\braket{\mathbf r|[\mathcal P_{\mathbf b_1}^{\pm},\mathcal P_{\mathbf b_2}^{\pm}]\mathcal P_{\Gamma}^{\pm}|\mathbf r}, \\
        &\mathcal C_{\pm}^{(sym)}(\mathbf r) =-\frac{1}{8\pi}\mathrm{Im}\bra{\mathbf r}\Big([\mathcal P_{\mathbf b_1}^{\pm},\mathcal P_{\mathbf b_2}^{\pm}]+[\mathcal P_{-\mathbf b_1}^{\pm},\mathcal P_{-\mathbf b_2}^{\pm}]-\nonumber \\
        & \hspace{1cm}-[\mathcal P_{-\mathbf b_1}^{\pm},\mathcal P_{\mathbf b_2}^{\pm}]-[\mathcal P_{\mathbf b_1}^{\pm},\mathcal P_{-\mathbf b_2}^{\pm}]\Big)\mathcal P_{\Gamma}^{\pm}\ket{\mathbf r},\label{eq:ilcm_spin_sym}
      \end{align}
      where $\mathcal P_{\Gamma}^{\pm}=\sum_{\lambda:\mathrm{sign}(s_{\lambda})=\pm}\ket{\phi_{\lambda\Gamma}}\bra{\phi_{\lambda\Gamma}}$, so that $\mathcal P_{\Gamma}=\mathcal P_{\Gamma}^++\mathcal P_{\Gamma}^-$. In particular, the covariant derivative is approximated by the forward finite difference formula in Eq.~\ref{eq:ilcm_spin} and by the symmetric finite difference in Eq.~\ref{eq:ilcm_spin_sym}. Finally, the PBC LSCM can be defined as:
      \begin{eqnarray}
        \nu(\mathbf r)=\frac{\mathcal C_+(\mathbf r)-\mathcal{C}_-(\mathbf r)}{2}\mod 2.
      \end{eqnarray}

    \subsection{Local PBC \texorpdfstring{$\mathbb{Z}_2$}{} marker based on time-reversal symmetry}
      Within PBCs, given a set of $J$ composite bands $\ket{\psi_{n\mathbf k}}$, the $m$-th WF in the unit cell labelled by the lattice vector $\mathbf R$ can be defined as:
        \begin{eqnarray}\label{eq:wannierfunctions_def}
          \ket{w_m(\mathbf R)}=\frac{A}{(2\pi)^2}\int_{BZ}d\mathbf k\,\, e^{-i\mathbf k\cdot\mathbf R}\sum_{n=1}^J U_{nm}^{(\mathbf k)}\ket{\psi_{n\mathbf k}}
        \end{eqnarray}
      where $U_{nm}^{(\mathbf k)}$ is a unitary rotation that is optimized to ensure the smoothness of the gauge. If such smooth and periodic gauge exists (that is, if the Chern number $C=0$) then the WFs are exponentially localized functions in real space~\cite{marzari_prl_2007}. As discussed in Sec.~\ref{sec:z2obc_b}, we compute the WFs of the system via projection onto trial states $\ket{g_n}$. In particular, the trial projections should break TR symmetry (to avoid the topological obstruction in the topological phase) and minimize the spillage between the projectors $\mathcal P$ and $\mathcal P_{\Theta}$, as defined in Eq.~\ref{eq:spillage_projectors}. Hence, we compute qWFs and split them in two subspaces mapped onto each other by TR symmetry, which allows computing the PBC individual local Chern markers $\mathcal C_{1,2}(\mathbf r)$ through Eq.~\ref{eq:ilcm_spin_sym}. Since the Hilbert space spanned by qWFs will be composed by pairs of states that are quasi-TR symmetric, we need to consider only one of these states when building the projectors $\mathcal P_{1,2}$. Finally, the PBC LZ2M can be defined as:
      \begin{eqnarray}
        \Delta(\mathbf r)=\frac{\mathcal C_1(\mathbf r)-\mathcal{C}_2(\mathbf r)}{2}\mod 2.
      \end{eqnarray}

  \section{Methods}\label{sec:methods}

    \subsection{Kane-Mele model}
      We validate our approach through numerical simulations on the Kane-Mele model~\cite{kane_quantum_2005,kane_$z_2$_2005}, a tight-binding model of spinful electrons hopping on a honeycomb lattice, described by the Hamiltonian:
      \begin{align}\label{eq:kanemelemodel}
        \mathcal H = \Delta &\sum_{i}(-1)^{\tau_i}c_{i}^{\dagger}c_{i} + t \sum_{\langle ij\rangle}c_i^{\dagger}c_j+i\lambda_{SO}\sum_{\langle\langle ij\rangle\rangle} \nu_{ij}c_i^{\dagger}\sigma_zc_j \nonumber\\&+ i\lambda_R\sum_{\langle ij\rangle}c_i^{\dagger}({\mathbf e}_{\langle ij\rangle}\cdot\boldsymbol\sigma)c_j + \mathrm{h.c.}
      \end{align}
      where $c_i^{\dagger}=(c_{i\uparrow}^{\dagger}, c_{i\downarrow}^{\dagger})$ and sums on spin indices are implied, with the convention that if no spin matrices appear, they are contracted over the identity. In Equation~\ref{eq:kanemelemodel}, $t$ is the nearest-neighbor hopping amplitude, and $\Delta$ is a staggered on-site potential, depending on the sublattice identified by $\tau_i\in\{0,1\}$. In the following, we will set $t=1$. The parameter $\lambda_{SO}$ is the intensity of the diagonal spin-orbit coupling, introduced as a complex hopping between second nearest-neighbors, where $\nu_{ij}=\mathrm{sign}(\mathbf d_1\times\mathbf d_2)_z$ accounts for the direction of the hopping and $\mathbf d_{1,2}$ are unit vectors connecting the site $i$ to its second nearest-neighbor $j$. Finally, $\lambda_R$ is the amplitude of the Rashba term, which couples the two spin sectors and breaks $S_z$ symmetry. Here $\boldsymbol \sigma=(\sigma_x, \sigma_y,\sigma_z)$ is the vector of Pauli matrices, and ${\mathbf e}_{\langle ij\rangle}=\mathbf{d}_{\langle ij\rangle}\times \hat{\mathbf z}$ where $\mathbf{d}_{\langle ij\rangle}$ is the unit vector along the direction connecting site $i$ to site $j$.

    \subsection{Smearing}
      When dealing with non-homogeneous systems, such as superlattices, the presence of metallic interfaces may affect the convergence of the topological marker. Hence, we introduce smearing similarly to what has been done in Ref.~\cite{marrazzo_prb_2017} to study the intrinsic geometrical part of the local anomalous Hall conductivity in metals. However, since the states appearing in the projectors $\mathcal P_{\mathbf b_j}^{\sigma}$ are linear combinations of the eigenstates of the Hamiltonian, we cannot simply use a Fermi-Dirac distribution as previously done for QAHIs in Refs.~\cite{marrazzo_prb_2017,pbclcm}. Hence, we introduce a weight $c_{n\mathbf b_j}^{\sigma}$ for each state $\ket{\tilde v_{n\mathbf b_j}^{\sigma}}$ (the parallel transported vectors of each subspace, so $\sigma=\pm$ or $\sigma=1,2$) that measures its spillage with the ground-state projector $\mathcal P_{\Gamma}$:
      \begin{align}\label{eq:spillage}
        c_{n\mathbf b_j}^{\sigma}&=\mathrm{Tr}\big\{ \mathcal P_{\Gamma}| \tilde v_{n\mathbf b_j}^{\sigma}\rangle\langle\tilde v_{n\mathbf b_j}^{\sigma}| \big\}\nonumber\\ &=\sum_m f(\epsilon_m, T_s, \mu)|\langle u_{m\Gamma}|\tilde v_{n\mathbf b_j}^{\sigma}\rangle|^2,
      \end{align}
      where $f(\epsilon_m, T_s, \mu)$ is the Fermi-Dirac distribution at smearing temperature $T_s$ and with chemical potential $\mu$, evaluated for the $m$-th eigenstate of the Hamiltonian $\mathcal H_{\Gamma}\ket{u_{m\Gamma}}=\epsilon_m\ket{u_{m\Gamma}}$. We further improve the convergence by imposing a cutoff $f_c$ on the Fermi-Dirac distribution to discard the empty states with very small occupations $f(\epsilon_m,T_s,\mu)<f_c$, where we set $f_c=0.1$. Finally, the projectors with smearing can be written as
      \begin{eqnarray}
        \mathcal P_{\mathbf b_j}^{\sigma}=\sum_n c_{n\mathbf b_j}^{\sigma}|\tilde v_{n\mathbf b_j}^{\sigma}\rangle\langle\tilde v_{n\mathbf b_j}^{\sigma}|.
      \end{eqnarray}

  \section{Numerical results and discussion}\label{sec:results}
    \subsection{Choosing the trial projections for the LZ2M}\label{sec:trial_projections}
\begin{figure}[h!]
        \centering
        \includegraphics[width=1\linewidth]{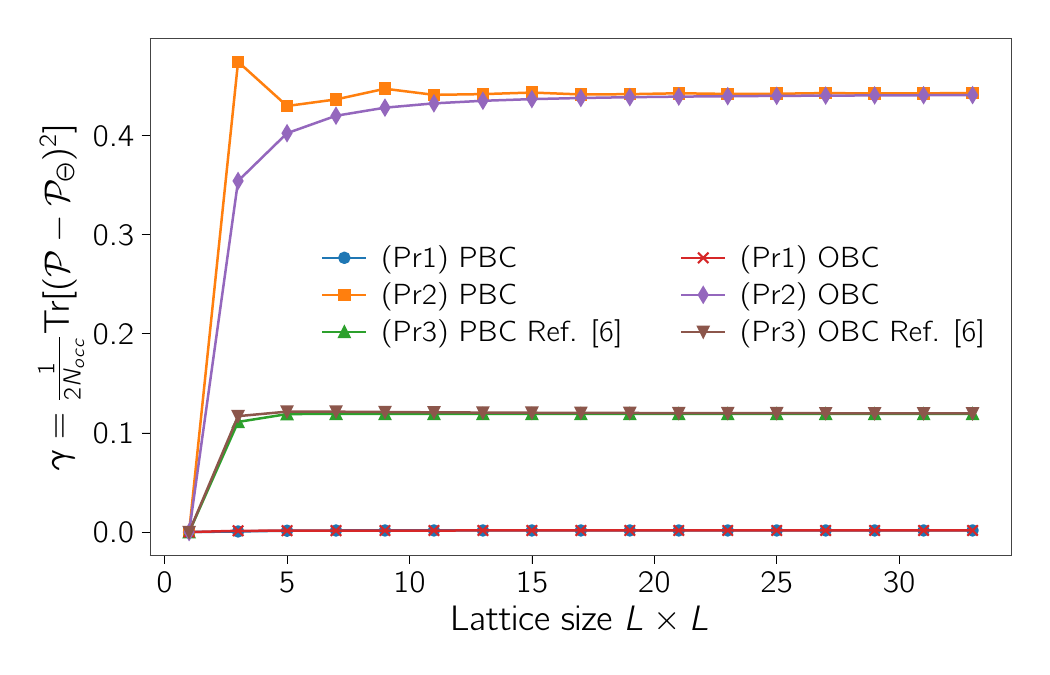}
        \caption{Spillage between the ground-state projector $\mathcal P$ and $\mathcal P_{\Theta}=\mathcal P_1+\Theta\mathcal P_1\Theta^{-1}$, where $\mathcal P_1$ is constructed with half of the quasi Wannier functions, as a function of the supercell size for both open (OBC) and periodic (PBC) boundary conditions, and for different choices of trial functions as discussed in the text.}
        \label{fig:1_1}
      \end{figure}

      The topological obstruction arising when $C\neq0$ manifests in the overlap matrix between projected states (the $S$ matrix of Eq.~\ref{eq:wf_projection}), that becomes singular somewhere in the BZ~\cite{topological_obstruction_chern}, prohibiting the construction of MLWFs.
      As discussed in Sec.~\ref{sec:z2obc_b}, in order to have well-defined local $\mathbb Z_2$ topological invariants based on TR symmetry, we split the occupied manifold in two TR-conjugate sub-manifolds and select trial projection orbitals $\ket{g_n}$ that break TR symmetry and minimize the spillage $\gamma$ between $\mathcal P_{\Theta}=\mathcal P_1+\Theta\mathcal P_1\Theta^{-1}$ and the ground-state projector $\mathcal P$. If we were to choose trial projections localized on just one site of the cell, the information on the other basis site would be missing since TR symmetry acts only on spin in real space, resulting in a large value of $\gamma$. Nonetheless, the resulting WFs can still have contributions from all the lattice sites, even though the trial projections are localized on different sites~\cite{wannier_z2_dimensions_soluyanov}. To minimize the spillage, we observe that it is convenient to choose initial projections with contributions from all sites. In the Kane-Mele model, this can be obtained, for instance, with these two choices:
      \begin{align}
        &(\mathrm{Pr1})\begin{cases}\label{eq:trb_1}
          \ket{g_1(\mathbf R)}=\frac{1}{\sqrt 2}(\ket{\mathbf R, A, +}+\ket{\mathbf R, B, +})\\
          \ket{g_2(\mathbf R)}=\frac{1}{\sqrt 2}(\ket{\mathbf R, A, -}-\ket{\mathbf R, B, -})
        \end{cases} \\
        &(\mathrm{Pr2})\begin{cases}\label{eq:trb_2}
          \ket{g_1(\mathbf R)}=\frac{1}{\sqrt 2}(\ket{\mathbf R, A, +}+\ket{\mathbf R, B, -})\\
          \ket{g_2(\mathbf R)}=\frac{1}{\sqrt 2}(\ket{\mathbf R, A, -}+\ket{\mathbf R, B, +})
        \end{cases}
      \end{align}
      where $\ket{\mathbf R, A, \sigma}$ is a normalized $\delta$-like orbital with spin $\sigma$ centered on site $A$ in the primitive cell defined by $\mathbf R$. Here we use the lattice vectors $\mathbf R$ of the pristine system defined in the primitive cell, as opposed to the lattice vectors of the supercell $\mathbf{\tilde R}$. Indeed, the supercell on which the boundary conditions are imposed is constructed through repetitions of its pristine primitive cell. Then, we can choose the trial projections just in the primitive cell and replicate the choice across the whole supercell, even in the disordered case. The spillage between the projectors $\mathcal P$ and $\mathcal P_{\Theta}$ can then be measured as a function of the linear size of the system $L$. The results we obtain are reported in Fig.~\ref{fig:1_1}, where we compare the spillage~\footnotemark[1] of the projections defined in Eqs.~\ref{eq:trb_1} and \ref{eq:trb_2} with the choice made in Ref.~\cite{soluyanov_wannier_z2_2011}, that is
      \begin{eqnarray}
        (\mathrm{Pr3})\begin{cases}\label{eq:vandb}
          \ket{g_1(\mathbf R)}=\frac{1}{\sqrt 2}(\ket{\mathbf R, A, +}+\ket{\mathbf R, A, -})\\
          \ket{g_2(\mathbf R)}=\frac{1}{\sqrt 2}(\ket{\mathbf R, B, +}-\ket{\mathbf R, B, -})
        \end{cases}.
      \end{eqnarray}
      The spillage between the projections (Pr3) defined in Eq.~\ref{eq:vandb} is small compared to other choices, we but since the trial functions are localized on a single site, the marker we can compute with those functions will not work. The projections (Pr2) [Eq.~\ref{eq:trb_2}] are localized on both sites but are characterized by a large spillage, so we can expect a poor convergence of the marker with this choice. Overall, the best choice is (Pr1) [Eq.~\ref{eq:trb_1}] and it will be used from now on when referring to the LZ2Ms: These orbitals mix all sites resulting in the smallest spillage. In general, to identify suitable initial projections, one could first compute the MLWFs and then mix them to minimize the spillage.
      When smearing is introduced in the calculation of the LZ2Ms, additional trial projections must be chosen to account for the partly occupied states. This case is analogous to that of a metal, where WFs can be computed via a ``disentanglement'' procedure~\cite{souza_disentanglement_prb_2001}. However, it looks unclear how the additional projections should be chosen, and how the procedure should be carried out to ensure the requirements described before. Indeed, one would have to localize the trial functions in an arbitrary unit cell in such a way as to preserve the exponential localization of the projectors and to make the marker independent with respect to the specific cell the trial functions are localized in. Hence, the construction of a LZ2M with smearing is left to further investigation.

    \subsection{Convergence}\label{sec:convergence}
\begin{figure}[h!]
        \centering
        \includegraphics[width=1\linewidth]{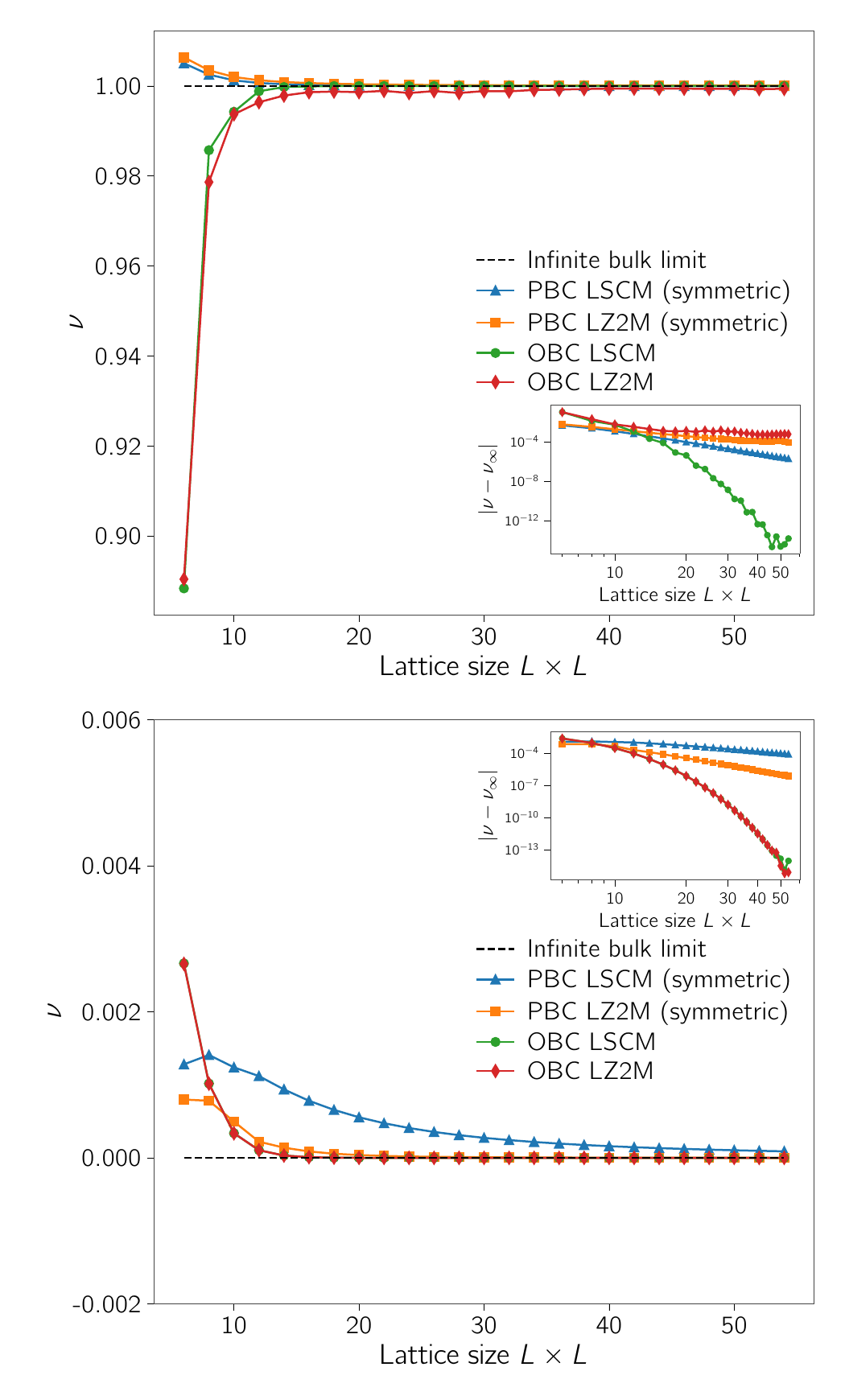}
        \caption{Convergence of the local topological markers in periodic (PBC) and open boundary conditions (OBC) for the topological (top) and trivial (bottom) phases. The convergence of the local spin-Chern marker (LSCM) is exponential in OBCs, and polynomial in PBCs. The local $\mathbb Z_2$ marker (LZ2M) convergence is always polynomial except for the trivial phase in OBCs where it is exponential.}
        \label{fig:2_1}
      \end{figure}

\begin{figure}[h!]
        \centering
        \includegraphics[width=1\linewidth]{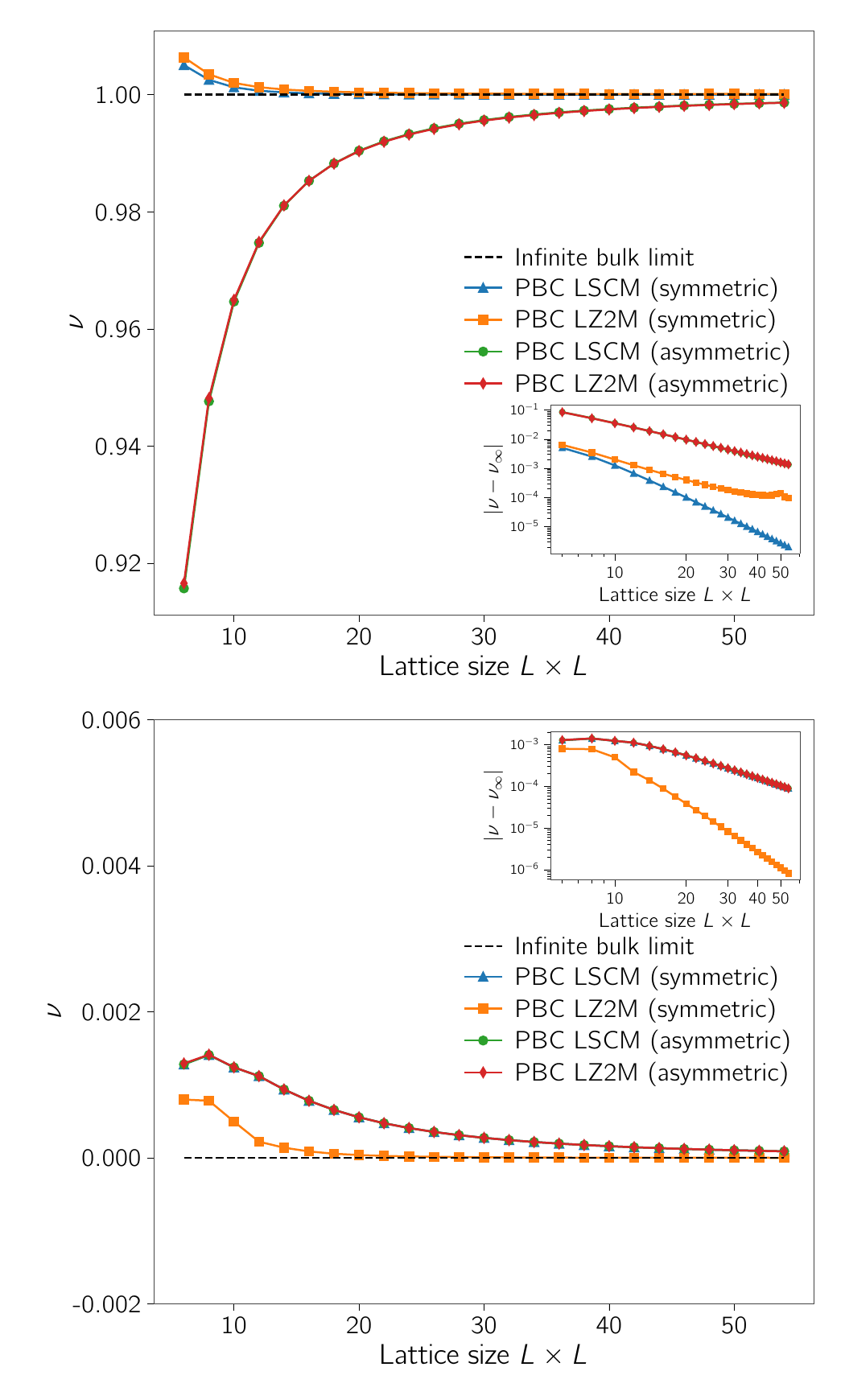}
        \caption{Convergence of the symmetric and asymmetric local spin-Chern markers (LSCM) and local $\mathbb Z_2$ markers (LZ2M) in periodic boundary conditions (PBC) for the topological (top) and trivial (bottom) phases. The symmetric formulas converge faster than the asymmetric ones, and the convergence of all markers is polynomial.}
        \label{fig:2_2}
      \end{figure}

\begin{figure}[h!]
        \centering
        \includegraphics[width=1\linewidth]{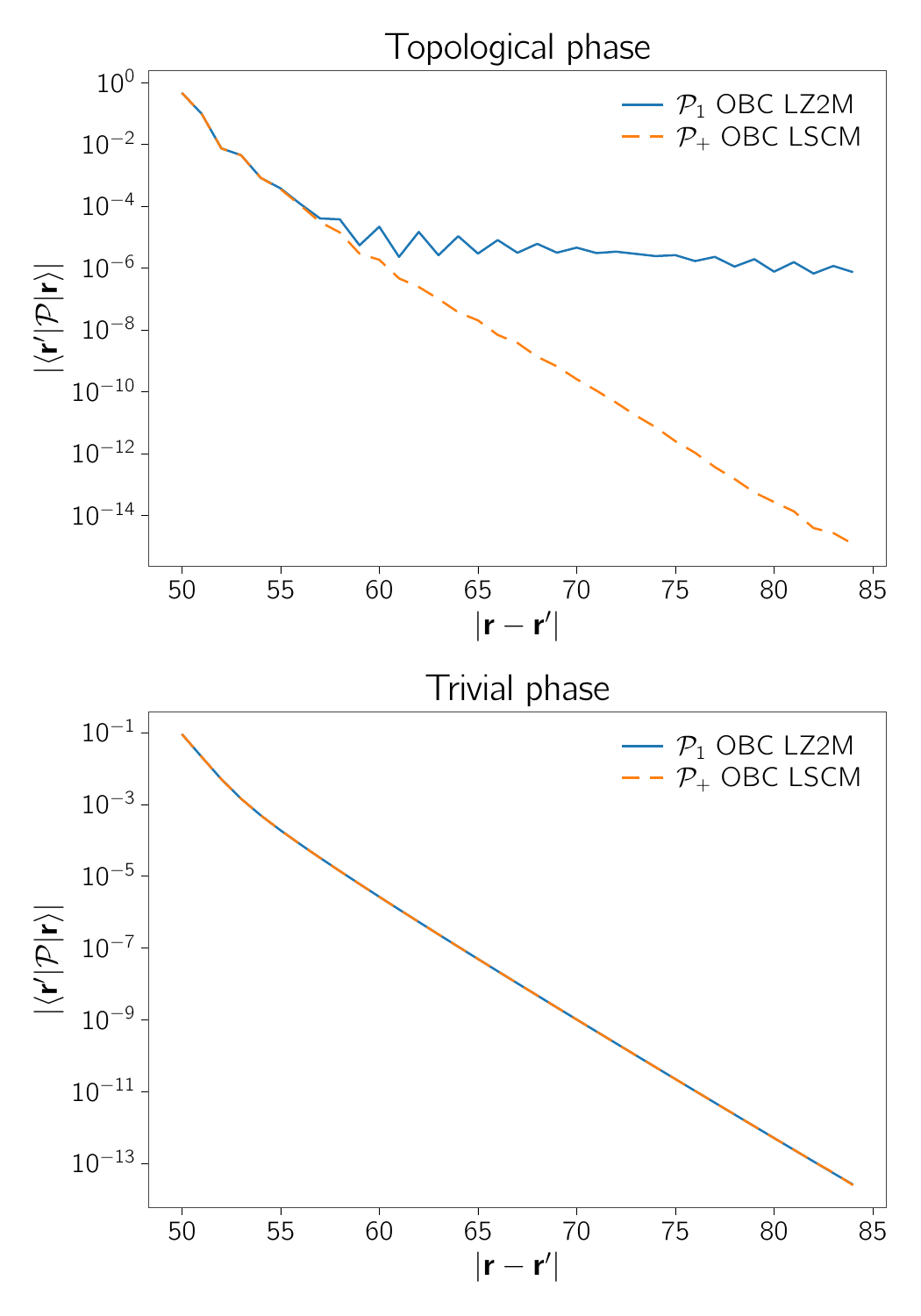}
        \caption{Matrix elements of the projectors $\mathcal P_+$ used in the local spin Chern marker (LSCM) and $\mathcal P_1$ of the local $\mathbb Z_2$ marker (LZ2M) inside the bulk as a function of the distance between lattice sites in the topological ($\nu=1$, top) and trivial ($\nu=0$, bottom) phases in open boundary conditions (OBC). The semilogarithmic scale highlights the exponential localization of the projectors, while the LZ2M projectors are exponentially localized at short distances and their tail decays more slowly in the topological phase. Nonetheless, $\mathcal P_{1,2}$ are localized enough to return the correct topological invariant of the system. In the trivial phase, both projectors are exponentially localized.}
        \label{fig:2_3}
      \end{figure}

      In Figure~\ref{fig:2_1}, we show the convergence of the LSCMs and LZ2Ms, evaluated on a unit cell (two sites) in the center of the supercell, as a function of the supercell linear size $L$ in both OBCs and PBCs. For the PBC markers, we report in Fig.~\ref{fig:2_1} the convergence of the symmetric formulations only, as they converge faster than the asymmetric ones. The comparison between the PBC asymmetric and symmetric formulas is reported in Fig.~\ref{fig:2_2}, that shows clearly the faster convergence of the symmetric formulations. The convergence of the PBC markers is polynomial in both the trivial and the topological phases since the error is dominated by the approximation of the derivative with finite differences in the single-point limit~\cite{pbclcm}. Within OBCs, we can see that the convergence of the LSCM is always exponential (due to the exponential localization of the projectors $P_{\pm}$~\cite{prodan_2009}), while the LZ2M converges exponentially in the trivial phase but polynomially in the topological one. This different behavior is due to the fact that, in the topological phase, the projector is exponentially localized only at short distances: Here, the projector is computed by first obtaining qWFs and then taking half of them. This can be clearly seen in Fig.~\ref{fig:2_3}, that shows, for the topological phase, the decay of the matrix elements of the projectors used in the OBC LSCM and OBC LZ2M in the bulk of a $100\times30$ crystallite with 6000 sites. We could not determine unambiguously whether this long tail is exponential or polynomial, that would require studying much larger systems.
      In the trivial phase, the projectors are always exponentially localized, resulting in an exponential convergence of the markers. We note that the LZ2M converges to the expected value despite its projectors not being fully exponentially localized. Indeed, since the invariant is a topological and a ground-state quantity, we only need the information about the local electronic structure to retrieve its local value in real space. For this reason, we argue that the projectors need to be exponentially localized only at short distances, while the tail behavior determines other aspects such as the convergence of the formula with the system size.
      We attribute the slower decay of the tails of $\mathcal P_1$ to be a side effect of minimizing the spillage $\gamma$ [Eq.~\ref{eq:spillage_projectors}] between the projectors $\mathcal P$ and $\mathcal P_{\Theta}$. Because of the topological obstruction, we cannot obtain TR symmetric and exponentially localized WFs. In our procedure, we require that the sum of the projector computed with half qWFs and their TR partners is as close as possible to the initial projector, still performing a minimal breaking of TR symmetry. That, in turn, results in an almost TR symmetric gauge that leads to qWFs that cannot be truly exponentially localized because of the topological obstruction. In other words, we are looking for a compromise between the breaking of TR symmetry and the exponential localization of WFs that allows us to probe the local $\mathbb Z_2$ topology in real space. As a result, the spread [Eq.~\ref{eq:spread}] of qWFs displays a very slow divergence with the system size, and the long-tail behavior of $\mathcal P_1$ could be attributed to the fact that the qWFs we obtain are not truly exponentially localized.
      

    \subsection{Disordered systems}
\begin{figure}[ht]
        \centering
        \includegraphics[width=1\linewidth]{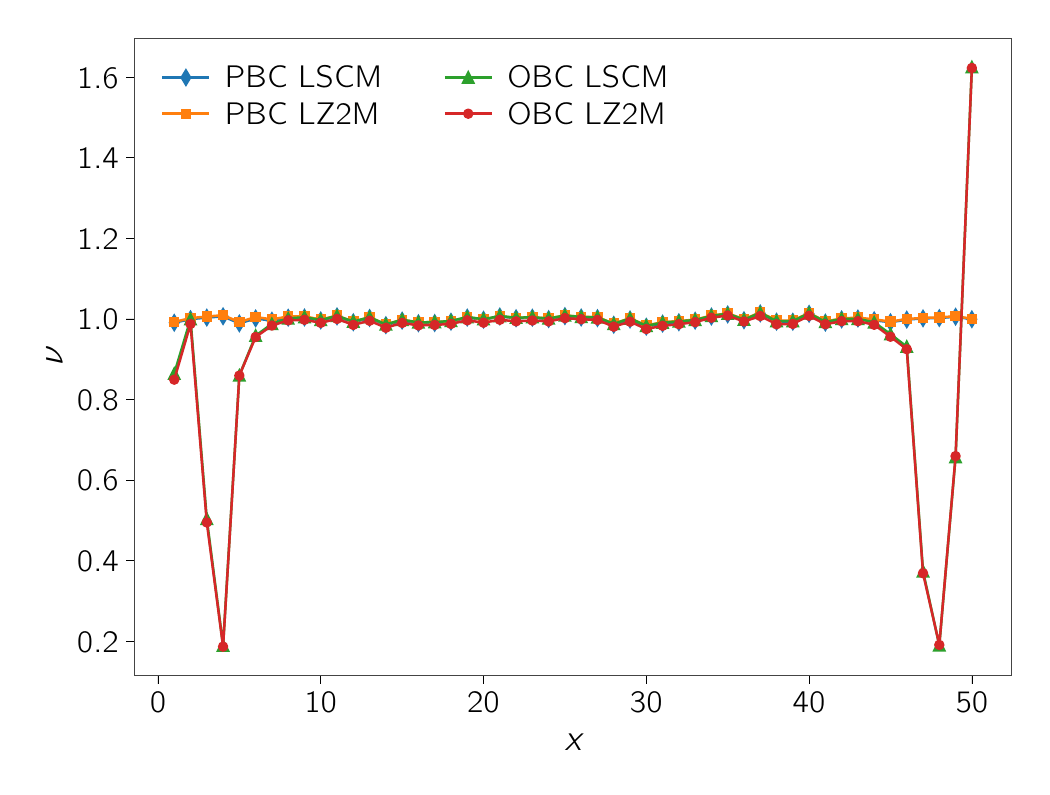}
        \caption{Profile along the $x$ direction of the local spin-Chern (LSCM) and local $\mathbb Z_2$ (LZ2M) markers with both open (OBC) and periodic (PBC) boundary conditions for the Kane-Mele model in the topological phase and in presence of Anderson disorder ($W=2$). First, a macroscopic average is performed over a radius $R=3$ and then over the two sites of the pristine lattice. The topological phase persists for sufficiently small disorder amplitudes, as confirmed by all markers. The discontinuous behavior of the OBC markers near the edges is due to the presence of metallic edge states, which are absent in PBCs.}
        \label{fig:3_1}
      \end{figure}

      In Figure~\ref{fig:3_1}, we show that both the LSCM and LZ2M are capable of charting the local topology also in presence disorder. In particular, we consider Anderson disorder through a random on-site potential uniformly distributed in the interval $[-W/2,W/2]$, where $W$ is the disorder amplitude~\cite{anderson_disorder_1958}. For weak disorder the topology of the system is expected to survive, hence we consider a Kane-Mele model in its topological phase ($\Delta/\lambda_{SO}=0.5$, $\lambda_R/\lambda_{SO}=1$) with $W=2$, for which the single-point spin-Chern number~\cite{favata_es_2023} predicts a topological phase. As shown in Fig.~\ref{fig:3_1}, the PBC markers display the expected local topology aside from small fluctuations due to the disordered environment. To account for the lack of periodicity, a macroscopic average on a radius $R=3$ (in units of the lattice parameter) has been employed. Specifically, the absence of boundaries of the supercell is evident when comparing the PBC markers with the OBC ones. In fact, despite showing the same behavior inside the bulk of the system, near the boundary the PBC markers are continuous while the OBC ones are not, due to the presence of metallic edge states. Within OBCs the individual LCMs $\mathfrak C_{\pm}(\mathbf r)$ are such that their trace over the whole sample vanishes, so the edge states contribution to the marker should compensate the non-trivial bulk topology. When computing the LSCM and LZ2M, since they are defined only modulo $2$, the divergence of the individual LCMs results in a discontinuous behavior of the $\mathbb Z_2$ markers.
    
    \subsection{Heterojunctions and superlattices}
\begin{figure*}[ht]
        \centering
        \includegraphics[width=1\linewidth]{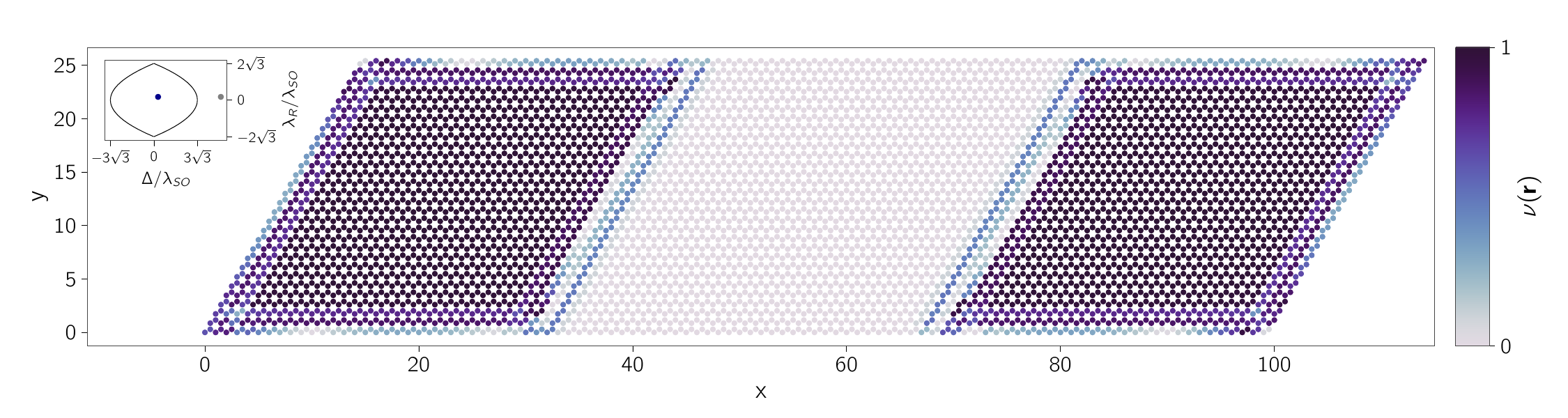}
        \caption{Local spin-Chern marker for a $6000$-site heterojunction of the Kane-Mele model made of topological and trivial regions in open boundary conditions. The left and right regions are topological ($\nu=1$) while the center is trivial ($\nu=0$). One-dimensional metallic edge states surround the topological regions, separating different topological phases. The inset displays the model parameters used for the trivial (gray) and topological (blue) regions.}
        \label{fig:4_1}
      \end{figure*}

\begin{figure*}[ht]
        \centering
        \includegraphics[width=1\linewidth]{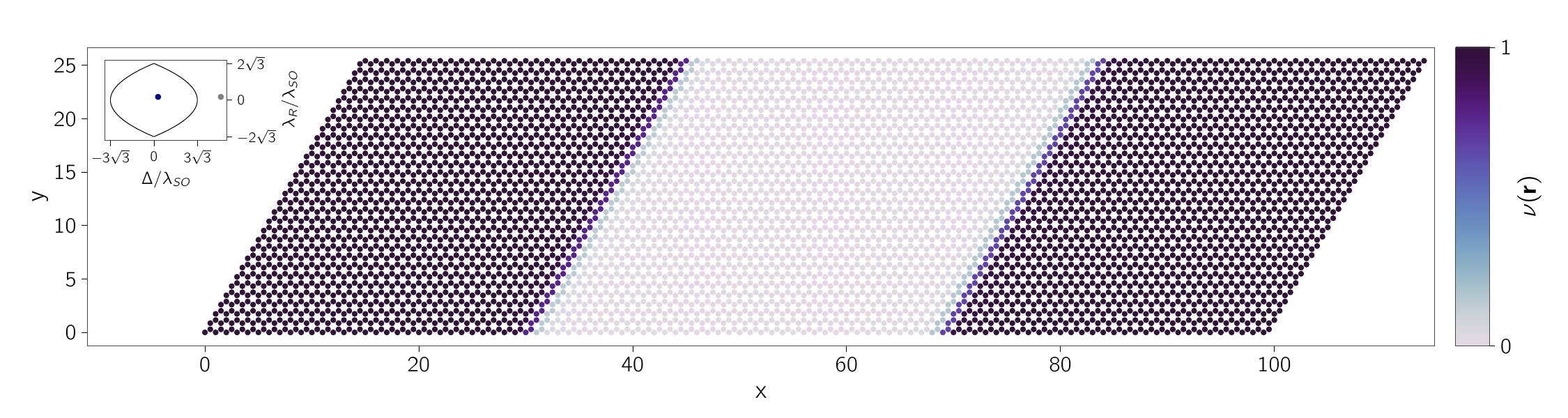}
        \caption{Local spin-Chern marker for a $6000$-site superlattice of the Kane-Mele model made of topological and trivial stripes in periodic boundary conditions. The left and right regions are topological ($\nu=1$) while the center is trivial ($\nu=0$), and one-dimensional metallic channels separate the regions with different $\mathbb Z_2$ invariants. The inset displays the model parameters used for the trivial (gray) and topological (blue) regions.}
        \label{fig:4_2}
      \end{figure*}

      Last, we validate and compare the performance of the local markers on trivial/topological heterojunctions and superlattices. We compute the LSCM in both OBCs [Fig.~\ref{fig:4_1}] and PBCs [Fig.~\ref{fig:4_2}] for a $6000$-site supercell made of alternating topological ($\Delta/\lambda_{SO}=0.5$) and trivial ($\Delta/\lambda_{SO}=8$) regions. Both the LSCM and LZ2M are able to chart the local topology, however, since the LSCM demonstrates a better convergence and performance with respect to the LZ2M we use only the former to discuss the numerical results for inhomogeneous systems. The presence of metallic edge states is highlighted in OBCs both at the edge of the supercell and at the interface between the subsystems, while in PBCs they appear only at the interface, as the superlattice is periodic in both directions. In the calculation of the PBC marker, we set the smearing temperature $T_s=0.05$ to improve convergence. Within OBCs, the Bianco-Resta LCM~\cite{bianco_prb_2011} vanishes when the trace is taken over the entire sample, and the topological marker evaluated at the metallic edges (where the Chern number is ill-defined) diverges. For QSHIs, the same holds for the local individual Chern markers $\mathfrak{C}_{\pm}(\mathbf{r})$. Here, since the local spin-Chern marker is well-defined only modulo $2$, the divergence of $\mathfrak{C}_{\pm}(\mathbf{r})$ results in the discontinuity observed at the interfaces between regions with different topology.


  \section{Summary and Conclusions}\label{sec:conclusions}
    We developed a framework to probe the local $\mathbb Z_2$ topology of 2D TR-symmetric systems in real space for both finite samples in OBCs and extended systems in PBCs, by introducing a number of topological markers. All markers are based on the fundamental idea that the occupied manifold can be split into two TR-conjugate subspaces, where a corresponding individual local Chern marker can then be calculated.
    In the first approach (LSCM), the separation is performed by diagonalizing the projected spin operator $\mathcal P S_z\mathcal P$, as its spectrum is generally made of two sectors of positive and negative eigenvalues separated by a gap, in the spirit of Prodan's spin-Chern number~\cite{prodan_2009}. Hence, the LSCM depends on the existence of such spectral gap, which is not an essential property of QSHIs but seems to be usually satisfied also in presence of rather strong Rashba SOC: We never observe such a gap closure for the $\mathcal P S_z\mathcal P$ spectrum in our simulations on the Kane-Mele model, at least as long as the Hamiltonian gap remains finite. However, as discussed in Ref.~\cite{layerlscm_natcomm_2024}, 3D strong topological insulators are characterized by a vanishing gap of $\mathcal PS\mathcal P$ for all choices of the spin operator $S$. Notably, the LSCM could in principle be used also for systems with a small breaking of TR symmetry, as in the TR-symmetry-broken quantum spin Hall effect~\cite{yang_prl_2011}, provided that the aforementioned spectral gap remains finite, since the construction does not explicitly make use of TR.

    Conceptually the extra condition required by Prodan's spin Chern number is not completely satisfactory, so we introduced another marker (LZ2M) that is entirely based on TR symmetry and does not make use of the spin operator. Crucially, Ref.~\cite{prodan_2010} remarked that exponentially localized projectors on each half of the manifold guarantee to yield an integer local Chern number. While this is guaranteed in the LSCM procedure by the existence of a spectral gap for $\mathcal P S_z\mathcal P$, this is not the case if TR is used to split the occupied manifold in two TR-conjugate halves.

    Hence, we find unitary rotations to obtain exponentially localized projectors onto the two subspaces, through the WF construction. However, QSHI do not admit a TR-symmetric smooth gauge, hence exponentially-localized WFs cannot be obtained as TR-conjugate couples. Still, the calculation of the $\mathbb Z_2$ invariant through the difference modulo two of individual Chern number requires a partition in two TR-conjugate subspaces. We solve the conundrum by constructing ``quasi'' WFs (qWFs) that minimize the spillage between the ground-state projector $\mathcal P$ and the sum of one TR-split projector and its TR-conjugate, i.e., $\mathcal P_1+\Theta\mathcal P_1\Theta^{-1}$. These qWFs are still exponentially localized at short distances but exhibit a slower asymptotic decay.

    Once the TR-conjugate subspaces and their corresponding well-localized projectors are found, we calculate local $\mathbb Z_2$ markers through local individual Chern markers on the two subspaces, both in OBCs ~\cite{bianco_prb_2011,marrazzo_prb_2017} and PBCs ~\cite{pbclcm}. Numerical simulations on the Kane-Mele model show that all these markers agree with each other and are able to probe the local topology also for disordered systems as well as topological/trivial heterojunctions and superlattices.

    An implementation of all our markers is available in the \texttt{StraWBerryPy} Python package~\cite{strawberrypy}, which is part of the Wannier function software ecosystem~\cite{WF_ecosys_2023} and is interfaced to popular tight-binding engines such as \texttt{TBmodels}~\cite{tbmodels,TB} and \texttt{PythTB}~\cite{pythtb}.

    The LSCM is computationally easier to implement and use, essentially avoiding the need of qWFs or any other localization procedure, so we suggest using that method in applications whenever possible. Otherwise, if the system of interest is such that the projected spin operator does not have a finite gap, the LZ2M approach can always be used (if TR symmetry holds). We emphasize that the construction of the LZ2Ms is more delicate, as an accurate choice of the initial projection orbitals is needed to obtain well-behaved topological markers. The specific projections we selected in our numerical experiments might be improved, and the choice is not universal, depending on the details of the system and its topological phase.

    Finally, it is worth remarking that the domains of applications for the LSCM and LZ2M markers mostly overlap, but there are some cases covered only by one of the two approaches: The LSCM requires a finite gap in the projected spin operator, but could in principle be used also if TR symmetry is broken; on the contrary, the LZ2M approach necessitates TR symmetry, but can be used also if the projected spin operator is gapless.
 
    Our local markers are based on the ground-state electron distribution only, hence being very suited to large-scale \textit{ab initio} electronic structure simulations of non-crystalline systems, not only under Anderson disorder but also in presence of defects or interfaces, and for amorphous topological materials~\cite{corbae_EPL_2023,corbae_natmat_2023,fazzio_nanolett_2019} or quasi-crystals~\cite{huang_prl_2018,huang_prb_2018}; in all cases even if TR symmetry or the perpendicular component of the spin, $S_z$, are not conserved. Moreover, the construction of our markers could potentially also be applied to the study of topological phonons~\cite{prodan_phonons_prl2009} and mechanical-optical setups~\cite{mechanical_topology_science}, where the dynamical matrix plays a role similar to the electronic Hamiltonian.

    \section{ACKNOWLEDGMENTS}
    N.B. acknowledges a scholarship from the Department of Physics of the University of Trieste and Collegio Universitario ``Luciano Fonda''; A.M. acknowledges that this study was partially funded by the European Union - NextGenerationEU, through the ICSC -- Centro Nazionale di Ricerca in High Performance Computing, Big Data and Quantum Computing -- (CUP J93C22000540006, PNRR Investimento M4.C2.1.4).  The views and opinions expressed are solely those of the authors and do not necessarily reflect those of the European Union, nor can the European Union be held responsible for them.

  \bibliography{biblio}

\begin{thebibliography}{74}%
\makeatletter
\providecommand \@ifxundefined [1]{%
 \@ifx{#1\undefined}
}%
\providecommand \@ifnum [1]{%
 \ifnum #1\expandafter \@firstoftwo
 \else \expandafter \@secondoftwo
 \fi
}%
\providecommand \@ifx [1]{%
 \ifx #1\expandafter \@firstoftwo
 \else \expandafter \@secondoftwo
 \fi
}%
\providecommand \natexlab [1]{#1}%
\providecommand \enquote  [1]{``#1''}%
\providecommand \bibnamefont  [1]{#1}%
\providecommand \bibfnamefont [1]{#1}%
\providecommand \citenamefont [1]{#1}%
\providecommand \href@noop [0]{\@secondoftwo}%
\providecommand \href [0]{\begingroup \@sanitize@url \@href}%
\providecommand \@href[1]{\@@startlink{#1}\@@href}%
\providecommand \@@href[1]{\endgroup#1\@@endlink}%
\providecommand \@sanitize@url [0]{\catcode `\\12\catcode `\$12\catcode
  `\&12\catcode `\#12\catcode `\^12\catcode `\_12\catcode `\%12\relax}%
\providecommand \@@startlink[1]{}%
\providecommand \@@endlink[0]{}%
\providecommand \url  [0]{\begingroup\@sanitize@url \@url }%
\providecommand \@url [1]{\endgroup\@href {#1}{\urlprefix }}%
\providecommand \urlprefix  [0]{URL }%
\providecommand \Eprint [0]{\href }%
\providecommand \doibase [0]{https://doi.org/}%
\providecommand \selectlanguage [0]{\@gobble}%
\providecommand \bibinfo  [0]{\@secondoftwo}%
\providecommand \bibfield  [0]{\@secondoftwo}%
\providecommand \translation [1]{[#1]}%
\providecommand \BibitemOpen [0]{}%
\providecommand \bibitemStop [0]{}%
\providecommand \bibitemNoStop [0]{.\EOS\space}%
\providecommand \EOS [0]{\spacefactor3000\relax}%
\providecommand \BibitemShut  [1]{\csname bibitem#1\endcsname}%
\let\auto@bib@innerbib\@empty
\bibitem [{\citenamefont {Haldane}(1988)}]{haldane_1988}%
  \BibitemOpen
  \bibfield  {author} {\bibinfo {author} {\bibfnamefont {F.~D.~M.}\
  \bibnamefont {Haldane}},\ }\bibfield  {title} {\bibinfo {title} {Model for a
  {Q}uantum {H}all {E}ffect without {L}andau {L}evels: {C}ondensed-{M}atter
  {R}ealization of the ``{P}arity {A}nomaly''},\ }\href
  {https://doi.org/10.1103/PhysRevLett.61.2015} {\bibfield  {journal} {\bibinfo
   {journal} {Phys. Rev. Lett.}\ }\textbf {\bibinfo {volume} {61}},\ \bibinfo
  {pages} {2015} (\bibinfo {year} {1988})}\BibitemShut {NoStop}%
\bibitem [{\citenamefont {Kane}\ and\ \citenamefont
  {Mele}(2005{\natexlab{a}})}]{kane_$z_2$_2005}%
  \BibitemOpen
  \bibfield  {author} {\bibinfo {author} {\bibfnamefont {C.~L.}\ \bibnamefont
  {Kane}}\ and\ \bibinfo {author} {\bibfnamefont {E.~J.}\ \bibnamefont
  {Mele}},\ }\bibfield  {title} {\bibinfo {title} {Z$_{2}$ {T}opological
  {O}rder and the {Q}uantum {S}pin {H}all {E}ffect},\ }\href
  {https://doi.org/10.1103/PhysRevLett.95.146802} {\bibfield  {journal}
  {\bibinfo  {journal} {Phys. Rev. Lett.}\ }\textbf {\bibinfo {volume} {95}},\
  \bibinfo {pages} {146802} (\bibinfo {year} {2005}{\natexlab{a}})}\BibitemShut
  {NoStop}%
\bibitem [{\citenamefont {Kane}\ and\ \citenamefont
  {Mele}(2005{\natexlab{b}})}]{kane_quantum_2005}%
  \BibitemOpen
  \bibfield  {author} {\bibinfo {author} {\bibfnamefont {C.~L.}\ \bibnamefont
  {Kane}}\ and\ \bibinfo {author} {\bibfnamefont {E.~J.}\ \bibnamefont
  {Mele}},\ }\bibfield  {title} {\bibinfo {title} {Quantum {S}pin {H}all
  {E}ffect in {G}raphene},\ }\href
  {https://doi.org/10.1103/PhysRevLett.95.226801} {\bibfield  {journal}
  {\bibinfo  {journal} {Phys. Rev. Lett.}\ }\textbf {\bibinfo {volume} {95}},\
  \bibinfo {pages} {226801} (\bibinfo {year} {2005}{\natexlab{b}})}\BibitemShut
  {NoStop}%
\bibitem [{\citenamefont {Bernevig}\ \emph {et~al.}(2006)\citenamefont
  {Bernevig}, \citenamefont {Hughes},\ and\ \citenamefont
  {Zhang}}]{bhzmodel_science_2006}%
  \BibitemOpen
  \bibfield  {author} {\bibinfo {author} {\bibfnamefont {B.~A.}\ \bibnamefont
  {Bernevig}}, \bibinfo {author} {\bibfnamefont {T.~L.}\ \bibnamefont
  {Hughes}},\ and\ \bibinfo {author} {\bibfnamefont {S.-C.}\ \bibnamefont
  {Zhang}},\ }\bibfield  {title} {\bibinfo {title} {{Quantum Spin Hall Effect
  and Topological Phase Transition in HgTe Quantum Wells}},\ }\href
  {https://doi.org/10.1126/science.1133734} {\bibfield  {journal} {\bibinfo
  {journal} {Science}\ }\textbf {\bibinfo {volume} {314}},\ \bibinfo {pages}
  {1757} (\bibinfo {year} {2006})}\BibitemShut {NoStop}%
\bibitem [{\citenamefont {Thonhauser}\ and\ \citenamefont
  {Vanderbilt}(2006)}]{topological_obstruction_chern}%
  \BibitemOpen
  \bibfield  {author} {\bibinfo {author} {\bibfnamefont {T.}~\bibnamefont
  {Thonhauser}}\ and\ \bibinfo {author} {\bibfnamefont {D.}~\bibnamefont
  {Vanderbilt}},\ }\bibfield  {title} {\bibinfo {title}
  {Insulator/{C}hern-insulator transition in the {H}aldane model},\ }\href
  {https://doi.org/10.1103/PhysRevB.74.235111} {\bibfield  {journal} {\bibinfo
  {journal} {Phys. Rev. B}\ }\textbf {\bibinfo {volume} {74}},\ \bibinfo
  {pages} {235111} (\bibinfo {year} {2006})}\BibitemShut {NoStop}%
\bibitem [{\citenamefont {Soluyanov}\ and\ \citenamefont
  {Vanderbilt}(2011{\natexlab{a}})}]{soluyanov_wannier_z2_2011}%
  \BibitemOpen
  \bibfield  {author} {\bibinfo {author} {\bibfnamefont {A.~A.}\ \bibnamefont
  {Soluyanov}}\ and\ \bibinfo {author} {\bibfnamefont {D.}~\bibnamefont
  {Vanderbilt}},\ }\bibfield  {title} {\bibinfo {title} {Wannier representation
  of $\mathbb{Z}_{2}$ topological insulators},\ }\href
  {https://doi.org/10.1103/PhysRevB.83.035108} {\bibfield  {journal} {\bibinfo
  {journal} {Phys. Rev. B}\ }\textbf {\bibinfo {volume} {83}},\ \bibinfo
  {pages} {035108} (\bibinfo {year} {2011}{\natexlab{a}})}\BibitemShut
  {NoStop}%
\bibitem [{\citenamefont {Fu}\ and\ \citenamefont
  {Kane}(2006)}]{ku_kane_trpolarization}%
  \BibitemOpen
  \bibfield  {author} {\bibinfo {author} {\bibfnamefont {L.}~\bibnamefont
  {Fu}}\ and\ \bibinfo {author} {\bibfnamefont {C.~L.}\ \bibnamefont {Kane}},\
  }\bibfield  {title} {\bibinfo {title} {Time reversal polarization and a
  ${Z}_{2}$ adiabatic spin pump},\ }\href
  {https://doi.org/10.1103/PhysRevB.74.195312} {\bibfield  {journal} {\bibinfo
  {journal} {Phys. Rev. B}\ }\textbf {\bibinfo {volume} {74}},\ \bibinfo
  {pages} {195312} (\bibinfo {year} {2006})}\BibitemShut {NoStop}%
\bibitem [{\citenamefont {Fu}\ and\ \citenamefont {Kane}(2007)}]{fu_kane_inv}%
  \BibitemOpen
  \bibfield  {author} {\bibinfo {author} {\bibfnamefont {L.}~\bibnamefont
  {Fu}}\ and\ \bibinfo {author} {\bibfnamefont {C.~L.}\ \bibnamefont {Kane}},\
  }\bibfield  {title} {\bibinfo {title} {Topological insulators with inversion
  symmetry},\ }\href {https://doi.org/10.1103/PhysRevB.76.045302} {\bibfield
  {journal} {\bibinfo  {journal} {Phys. Rev. B}\ }\textbf {\bibinfo {volume}
  {76}},\ \bibinfo {pages} {045302} (\bibinfo {year} {2007})}\BibitemShut
  {NoStop}%
\bibitem [{\citenamefont {Yu}\ \emph {et~al.}(2011)\citenamefont {Yu},
  \citenamefont {Qi}, \citenamefont {Bernevig}, \citenamefont {Fang},\ and\
  \citenamefont {Dai}}]{yu_z2_wf_prb_2011}%
  \BibitemOpen
  \bibfield  {author} {\bibinfo {author} {\bibfnamefont {R.}~\bibnamefont
  {Yu}}, \bibinfo {author} {\bibfnamefont {X.~L.}\ \bibnamefont {Qi}}, \bibinfo
  {author} {\bibfnamefont {A.}~\bibnamefont {Bernevig}}, \bibinfo {author}
  {\bibfnamefont {Z.}~\bibnamefont {Fang}},\ and\ \bibinfo {author}
  {\bibfnamefont {X.}~\bibnamefont {Dai}},\ }\bibfield  {title} {\bibinfo
  {title} {Equivalent expression of $\mathbb{Z}_{2}$ topological invariant for
  band insulators using the non-{A}belian {B}erry connection},\ }\href
  {https://doi.org/10.1103/PhysRevB.84.075119} {\bibfield  {journal} {\bibinfo
  {journal} {Phys. Rev. B}\ }\textbf {\bibinfo {volume} {84}},\ \bibinfo
  {pages} {075119} (\bibinfo {year} {2011})}\BibitemShut {NoStop}%
\bibitem [{\citenamefont {Gresch}\ \emph {et~al.}(2017)\citenamefont {Gresch},
  \citenamefont {Aut\`es}, \citenamefont {Yazyev}, \citenamefont {Troyer},
  \citenamefont {Vanderbilt}, \citenamefont {Bernevig},\ and\ \citenamefont
  {Soluyanov}}]{z2pack}%
  \BibitemOpen
  \bibfield  {author} {\bibinfo {author} {\bibfnamefont {D.}~\bibnamefont
  {Gresch}}, \bibinfo {author} {\bibfnamefont {G.}~\bibnamefont {Aut\`es}},
  \bibinfo {author} {\bibfnamefont {O.~V.}\ \bibnamefont {Yazyev}}, \bibinfo
  {author} {\bibfnamefont {M.}~\bibnamefont {Troyer}}, \bibinfo {author}
  {\bibfnamefont {D.}~\bibnamefont {Vanderbilt}}, \bibinfo {author}
  {\bibfnamefont {B.~A.}\ \bibnamefont {Bernevig}},\ and\ \bibinfo {author}
  {\bibfnamefont {A.~A.}\ \bibnamefont {Soluyanov}},\ }\bibfield  {title}
  {\bibinfo {title} {Z2{P}ack: {N}umerical implementation of hybrid {W}annier
  centers for identifying topological materials},\ }\href
  {https://doi.org/10.1103/PhysRevB.95.075146} {\bibfield  {journal} {\bibinfo
  {journal} {Phys. Rev. B}\ }\textbf {\bibinfo {volume} {95}},\ \bibinfo
  {pages} {075146} (\bibinfo {year} {2017})}\BibitemShut {NoStop}%
\bibitem [{\citenamefont {Ceresoli}\ and\ \citenamefont
  {Resta}(2007)}]{ceresoli_sp_prb_2007}%
  \BibitemOpen
  \bibfield  {author} {\bibinfo {author} {\bibfnamefont {D.}~\bibnamefont
  {Ceresoli}}\ and\ \bibinfo {author} {\bibfnamefont {R.}~\bibnamefont
  {Resta}},\ }\bibfield  {title} {\bibinfo {title} {Orbital magnetization and
  {C}hern number in a supercell framework: {S}ingle $\mathbf{k}$-point
  formula},\ }\href {https://doi.org/10.1103/PhysRevB.76.012405} {\bibfield
  {journal} {\bibinfo  {journal} {Phys. Rev. B}\ }\textbf {\bibinfo {volume}
  {76}},\ \bibinfo {pages} {012405} (\bibinfo {year} {2007})}\BibitemShut
  {NoStop}%
\bibitem [{\citenamefont {Favata}\ and\ \citenamefont
  {Marrazzo}(2023)}]{favata_es_2023}%
  \BibitemOpen
  \bibfield  {author} {\bibinfo {author} {\bibfnamefont {R.}~\bibnamefont
  {Favata}}\ and\ \bibinfo {author} {\bibfnamefont {A.}~\bibnamefont
  {Marrazzo}},\ }\bibfield  {title} {\bibinfo {title} {Single-point spin
  {C}hern number in a supercell framework},\ }\href
  {https://doi.org/10.1088/2516-1075/acba6f} {\bibfield  {journal} {\bibinfo
  {journal} {Electronic Structure}\ }\textbf {\bibinfo {volume} {5}},\ \bibinfo
  {pages} {014005} (\bibinfo {year} {2023})}\BibitemShut {NoStop}%
\bibitem [{\citenamefont {Loring}\ and\ \citenamefont
  {Hastings}(2011)}]{Loring_2011}%
  \BibitemOpen
  \bibfield  {author} {\bibinfo {author} {\bibfnamefont {T.~A.}\ \bibnamefont
  {Loring}}\ and\ \bibinfo {author} {\bibfnamefont {M.~B.}\ \bibnamefont
  {Hastings}},\ }\bibfield  {title} {\bibinfo {title} {Disordered topological
  insulators via {$C^*$}-algebras},\ }\href
  {https://doi.org/10.1209/0295-5075/92/67004} {\bibfield  {journal} {\bibinfo
  {journal} {Europhysics Letters}\ }\textbf {\bibinfo {volume} {92}},\ \bibinfo
  {pages} {67004} (\bibinfo {year} {2011})}\BibitemShut {NoStop}%
\bibitem [{\citenamefont {Huang}\ and\ \citenamefont
  {Liu}(2018{\natexlab{a}})}]{huang_prb_2018}%
  \BibitemOpen
  \bibfield  {author} {\bibinfo {author} {\bibfnamefont {H.}~\bibnamefont
  {Huang}}\ and\ \bibinfo {author} {\bibfnamefont {F.}~\bibnamefont {Liu}},\
  }\bibfield  {title} {\bibinfo {title} {Theory of spin {B}ott index for
  quantum spin {H}all states in nonperiodic systems},\ }\href
  {https://doi.org/10.1103/PhysRevB.98.125130} {\bibfield  {journal} {\bibinfo
  {journal} {Phys. Rev. B}\ }\textbf {\bibinfo {volume} {98}},\ \bibinfo
  {pages} {125130} (\bibinfo {year} {2018}{\natexlab{a}})}\BibitemShut
  {NoStop}%
\bibitem [{\citenamefont {Mu\~noz{-}Segovia}\ \emph {et~al.}(2023)\citenamefont
  {Mu\~noz{-}Segovia}, \citenamefont {Corbae}, \citenamefont {Varjas},
  \citenamefont {Hellman}, \citenamefont {Griffin},\ and\ \citenamefont
  {Grushin}}]{grushin_prr_2023}%
  \BibitemOpen
  \bibfield  {author} {\bibinfo {author} {\bibfnamefont {D.}~\bibnamefont
  {Mu\~noz{-}Segovia}}, \bibinfo {author} {\bibfnamefont {P.}~\bibnamefont
  {Corbae}}, \bibinfo {author} {\bibfnamefont {D.}~\bibnamefont {Varjas}},
  \bibinfo {author} {\bibfnamefont {F.}~\bibnamefont {Hellman}}, \bibinfo
  {author} {\bibfnamefont {S.~M.}\ \bibnamefont {Griffin}},\ and\ \bibinfo
  {author} {\bibfnamefont {A.~G.}\ \bibnamefont {Grushin}},\ }\bibfield
  {title} {\bibinfo {title} {{Structural spillage: An efficient method to
  identify noncrystalline topological materials}},\ }\href
  {https://doi.org/10.1103/PhysRevResearch.5.L042011} {\bibfield  {journal}
  {\bibinfo  {journal} {Phys. Rev. Res.}\ }\textbf {\bibinfo {volume} {5}},\
  \bibinfo {pages} {L042011} (\bibinfo {year} {2023})}\BibitemShut {NoStop}%
\bibitem [{\citenamefont {Fulga}\ \emph {et~al.}(2012)\citenamefont {Fulga},
  \citenamefont {Hassler},\ and\ \citenamefont
  {Akhmerov}}]{PhysRevB.85.165409}%
  \BibitemOpen
  \bibfield  {author} {\bibinfo {author} {\bibfnamefont {I.~C.}\ \bibnamefont
  {Fulga}}, \bibinfo {author} {\bibfnamefont {F.}~\bibnamefont {Hassler}},\
  and\ \bibinfo {author} {\bibfnamefont {A.~R.}\ \bibnamefont {Akhmerov}},\
  }\bibfield  {title} {\bibinfo {title} {{Scattering theory of topological
  insulators and superconductors}},\ }\href
  {https://doi.org/10.1103/PhysRevB.85.165409} {\bibfield  {journal} {\bibinfo
  {journal} {Phys. Rev. B}\ }\textbf {\bibinfo {volume} {85}},\ \bibinfo
  {pages} {165409} (\bibinfo {year} {2012})}\BibitemShut {NoStop}%
\bibitem [{\citenamefont {Fulga}\ \emph {et~al.}(2011)\citenamefont {Fulga},
  \citenamefont {Hassler}, \citenamefont {Akhmerov},\ and\ \citenamefont
  {Beenakker}}]{PhysRevB.83.155429}%
  \BibitemOpen
  \bibfield  {author} {\bibinfo {author} {\bibfnamefont {I.~C.}\ \bibnamefont
  {Fulga}}, \bibinfo {author} {\bibfnamefont {F.}~\bibnamefont {Hassler}},
  \bibinfo {author} {\bibfnamefont {A.~R.}\ \bibnamefont {Akhmerov}},\ and\
  \bibinfo {author} {\bibfnamefont {C.~W.~J.}\ \bibnamefont {Beenakker}},\
  }\bibfield  {title} {\bibinfo {title} {{Scattering formula for the
  topological quantum number of a disordered multimode wire}},\ }\href
  {https://doi.org/10.1103/PhysRevB.83.155429} {\bibfield  {journal} {\bibinfo
  {journal} {Phys. Rev. B}\ }\textbf {\bibinfo {volume} {83}},\ \bibinfo
  {pages} {155429} (\bibinfo {year} {2011})}\BibitemShut {NoStop}%
\bibitem [{\citenamefont {Avron}\ \emph
  {et~al.}(1994{\natexlab{a}})\citenamefont {Avron}, \citenamefont {Seiler},\
  and\ \citenamefont {Simon}}]{Avron1994}%
  \BibitemOpen
  \bibfield  {author} {\bibinfo {author} {\bibfnamefont {J.~E.}\ \bibnamefont
  {Avron}}, \bibinfo {author} {\bibfnamefont {R.}~\bibnamefont {Seiler}},\ and\
  \bibinfo {author} {\bibfnamefont {B.}~\bibnamefont {Simon}},\ }\bibfield
  {title} {\bibinfo {title} {{Charge deficiency, charge transport and
  comparison of dimensions}},\ }\href {https://doi.org/10.1007/BF02102644}
  {\bibfield  {journal} {\bibinfo  {journal} {Communications in Mathematical
  Physics}\ }\textbf {\bibinfo {volume} {159}},\ \bibinfo {pages} {399}
  (\bibinfo {year} {1994}{\natexlab{a}})}\BibitemShut {NoStop}%
\bibitem [{\citenamefont {Avron}\ \emph
  {et~al.}(1994{\natexlab{b}})\citenamefont {Avron}, \citenamefont {Seiler},\
  and\ \citenamefont {Simon}}]{AVRON1994220}%
  \BibitemOpen
  \bibfield  {author} {\bibinfo {author} {\bibfnamefont {J.}~\bibnamefont
  {Avron}}, \bibinfo {author} {\bibfnamefont {R.}~\bibnamefont {Seiler}},\ and\
  \bibinfo {author} {\bibfnamefont {B.}~\bibnamefont {Simon}},\ }\bibfield
  {title} {\bibinfo {title} {The {I}ndex of a {P}air of {P}rojections},\ }\href
  {https://doi.org/10.1006/jfan.1994.1031} {\bibfield  {journal} {\bibinfo
  {journal} {Journal of Functional Analysis}\ }\textbf {\bibinfo {volume}
  {120}},\ \bibinfo {pages} {220} (\bibinfo {year}
  {1994}{\natexlab{b}})}\BibitemShut {NoStop}%
\bibitem [{\citenamefont {Katsura}\ and\ \citenamefont
  {Koma}(2018)}]{10.1063/1.5026964}%
  \BibitemOpen
  \bibfield  {author} {\bibinfo {author} {\bibfnamefont {H.}~\bibnamefont
  {Katsura}}\ and\ \bibinfo {author} {\bibfnamefont {T.}~\bibnamefont {Koma}},\
  }\bibfield  {title} {\bibinfo {title} {{The noncommutative index theorem and
  the periodic table for disordered topological insulators and
  superconductors}},\ }\href {https://doi.org/10.1063/1.5026964} {\bibfield
  {journal} {\bibinfo  {journal} {Journal of Mathematical Physics}\ }\textbf
  {\bibinfo {volume} {59}},\ \bibinfo {pages} {031903} (\bibinfo {year}
  {2018})}\BibitemShut {NoStop}%
\bibitem [{\citenamefont {Bianco}\ and\ \citenamefont
  {Resta}(2011)}]{bianco_prb_2011}%
  \BibitemOpen
  \bibfield  {author} {\bibinfo {author} {\bibfnamefont {R.}~\bibnamefont
  {Bianco}}\ and\ \bibinfo {author} {\bibfnamefont {R.}~\bibnamefont {Resta}},\
  }\bibfield  {title} {\bibinfo {title} {Mapping topological order in
  coordinate space},\ }\href {https://doi.org/10.1103/PhysRevB.84.241106}
  {\bibfield  {journal} {\bibinfo  {journal} {Phys. Rev. B}\ }\textbf {\bibinfo
  {volume} {84}},\ \bibinfo {pages} {241106(R)} (\bibinfo {year}
  {2011})}\BibitemShut {NoStop}%
\bibitem [{\citenamefont {Varjas}\ \emph {et~al.}(2020)\citenamefont {Varjas},
  \citenamefont {Fruchart}, \citenamefont {Akhmerov},\ and\ \citenamefont
  {Perez-Piskunow}}]{kpm_lcm_2020}%
  \BibitemOpen
  \bibfield  {author} {\bibinfo {author} {\bibfnamefont {D.}~\bibnamefont
  {Varjas}}, \bibinfo {author} {\bibfnamefont {M.}~\bibnamefont {Fruchart}},
  \bibinfo {author} {\bibfnamefont {A.~R.}\ \bibnamefont {Akhmerov}},\ and\
  \bibinfo {author} {\bibfnamefont {P.~M.}\ \bibnamefont {Perez-Piskunow}},\
  }\bibfield  {title} {\bibinfo {title} {Computation of topological phase
  diagram of disordered
  $\mathrm{Pb}_{1\ensuremath{-}x}\mathrm{Sn}_{x}\mathrm{Te}$ using the kernel
  polynomial method},\ }\href
  {https://doi.org/10.1103/PhysRevResearch.2.013229} {\bibfield  {journal}
  {\bibinfo  {journal} {Phys. Rev. Res.}\ }\textbf {\bibinfo {volume} {2}},\
  \bibinfo {pages} {013229} (\bibinfo {year} {2020})}\BibitemShut {NoStop}%
\bibitem [{\citenamefont {Gebert}\ \emph {et~al.}(2020)\citenamefont {Gebert},
  \citenamefont {Irsigler},\ and\ \citenamefont
  {Hofstetter}}]{confinedhoftadter_gebert_2020}%
  \BibitemOpen
  \bibfield  {author} {\bibinfo {author} {\bibfnamefont {U.}~\bibnamefont
  {Gebert}}, \bibinfo {author} {\bibfnamefont {B.}~\bibnamefont {Irsigler}},\
  and\ \bibinfo {author} {\bibfnamefont {W.}~\bibnamefont {Hofstetter}},\
  }\bibfield  {title} {\bibinfo {title} {Local {C}hern marker of smoothly
  confined {H}ofstadter fermions},\ }\href
  {https://doi.org/10.1103/PhysRevA.101.063606} {\bibfield  {journal} {\bibinfo
   {journal} {Phys. Rev. A}\ }\textbf {\bibinfo {volume} {101}},\ \bibinfo
  {pages} {063606} (\bibinfo {year} {2020})}\BibitemShut {NoStop}%
\bibitem [{\citenamefont {Ul\ifmmode~\check{c}\else \v{c}\fi{}akar}\ \emph
  {et~al.}(2020)\citenamefont {Ul\ifmmode~\check{c}\else \v{c}\fi{}akar},
  \citenamefont {Mravlje},\ and\ \citenamefont
  {Rejec}}]{disorderchern_prl_2020}%
  \BibitemOpen
  \bibfield  {author} {\bibinfo {author} {\bibfnamefont {L.}~\bibnamefont
  {Ul\ifmmode~\check{c}\else \v{c}\fi{}akar}}, \bibinfo {author} {\bibfnamefont
  {J.}~\bibnamefont {Mravlje}},\ and\ \bibinfo {author} {\bibfnamefont
  {T.~c.~v.}\ \bibnamefont {Rejec}},\ }\bibfield  {title} {\bibinfo {title}
  {{Kibble-Zurek Behavior in Disordered Chern Insulators}},\ }\href
  {https://doi.org/10.1103/PhysRevLett.125.216601} {\bibfield  {journal}
  {\bibinfo  {journal} {Phys. Rev. Lett.}\ }\textbf {\bibinfo {volume} {125}},\
  \bibinfo {pages} {216601} (\bibinfo {year} {2020})}\BibitemShut {NoStop}%
\bibitem [{\citenamefont {d'Ornellas}\ \emph {et~al.}(2022)\citenamefont
  {d'Ornellas}, \citenamefont {Barnett},\ and\ \citenamefont
  {Lee}}]{dornellas_prb_2022}%
  \BibitemOpen
  \bibfield  {author} {\bibinfo {author} {\bibfnamefont {P.}~\bibnamefont
  {d'Ornellas}}, \bibinfo {author} {\bibfnamefont {R.}~\bibnamefont
  {Barnett}},\ and\ \bibinfo {author} {\bibfnamefont {D.~K.~K.}\ \bibnamefont
  {Lee}},\ }\bibfield  {title} {\bibinfo {title} {Quantized bulk conductivity
  as a local {C}hern marker},\ }\href
  {https://doi.org/10.1103/PhysRevB.106.155124} {\bibfield  {journal} {\bibinfo
   {journal} {Phys. Rev. B}\ }\textbf {\bibinfo {volume} {106}},\ \bibinfo
  {pages} {155124} (\bibinfo {year} {2022})}\BibitemShut {NoStop}%
\bibitem [{\citenamefont {Mildner}\ \emph {et~al.}(2023)\citenamefont
  {Mildner}, \citenamefont {Caio}, \citenamefont {Möller}, \citenamefont
  {Cooper},\ and\ \citenamefont {Bhaseen}}]{mildner_2023}%
  \BibitemOpen
  \bibfield  {author} {\bibinfo {author} {\bibfnamefont {J.}~\bibnamefont
  {Mildner}}, \bibinfo {author} {\bibfnamefont {M.~D.}\ \bibnamefont {Caio}},
  \bibinfo {author} {\bibfnamefont {G.}~\bibnamefont {Möller}}, \bibinfo
  {author} {\bibfnamefont {N.~R.}\ \bibnamefont {Cooper}},\ and\ \bibinfo
  {author} {\bibfnamefont {M.~J.}\ \bibnamefont {Bhaseen}},\ }\href@noop {}
  {\bibinfo {title} {Topological {P}hase {T}ransitions in the {D}isordered
  {H}aldane {M}odel}} (\bibinfo {year} {2023}),\ \Eprint
  {https://arxiv.org/abs/2312.16689} {arXiv:2312.16689 [cond-mat.str-el]}
  \BibitemShut {NoStop}%
\bibitem [{\citenamefont {Marsal}\ \emph
  {et~al.}(2020{\natexlab{a}})\citenamefont {Marsal}, \citenamefont {Varjas},\
  and\ \citenamefont {Grushin}}]{grushin_amorphousti_2020}%
  \BibitemOpen
  \bibfield  {author} {\bibinfo {author} {\bibfnamefont {Q.}~\bibnamefont
  {Marsal}}, \bibinfo {author} {\bibfnamefont {D.}~\bibnamefont {Varjas}},\
  and\ \bibinfo {author} {\bibfnamefont {A.~G.}\ \bibnamefont {Grushin}},\
  }\bibfield  {title} {\bibinfo {title} {Topological {W}eaire-{T}horpe models
  of amorphous matter},\ }\href {https://doi.org/10.1073/pnas.2007384117}
  {\bibfield  {journal} {\bibinfo  {journal} {Proceedings of the National
  Academy of Sciences}\ }\textbf {\bibinfo {volume} {117}},\ \bibinfo {pages}
  {30260} (\bibinfo {year} {2020}{\natexlab{a}})}\BibitemShut {NoStop}%
\bibitem [{\citenamefont {Kohn}(1996)}]{kohn_prl_1996}%
  \BibitemOpen
  \bibfield  {author} {\bibinfo {author} {\bibfnamefont {W.}~\bibnamefont
  {Kohn}},\ }\bibfield  {title} {\bibinfo {title} {Density {F}unctional and
  {D}ensity {M}atrix {M}ethod {S}caling {L}inearly with the {N}umber of
  {A}toms},\ }\href {https://doi.org/10.1103/PhysRevLett.76.3168} {\bibfield
  {journal} {\bibinfo  {journal} {Phys. Rev. Lett.}\ }\textbf {\bibinfo
  {volume} {76}},\ \bibinfo {pages} {3168} (\bibinfo {year}
  {1996})}\BibitemShut {NoStop}%
\bibitem [{\citenamefont {Assun\c{c}\~ao}\ \emph {et~al.}(2024)\citenamefont
  {Assun\c{c}\~ao}, \citenamefont {Ferreira},\ and\ \citenamefont
  {Lewenkopf}}]{assuncao_lcmBHZ_2024}%
  \BibitemOpen
  \bibfield  {author} {\bibinfo {author} {\bibfnamefont {B.~D.}\ \bibnamefont
  {Assun\c{c}\~ao}}, \bibinfo {author} {\bibfnamefont {G.~J.}\ \bibnamefont
  {Ferreira}},\ and\ \bibinfo {author} {\bibfnamefont {C.~H.}\ \bibnamefont
  {Lewenkopf}},\ }\bibfield  {title} {\bibinfo {title} {Phase transitions and
  scale invariance in topological {A}nderson insulators},\ }\href
  {https://doi.org/10.1103/PhysRevB.109.L201102} {\bibfield  {journal}
  {\bibinfo  {journal} {Phys. Rev. B}\ }\textbf {\bibinfo {volume} {109}},\
  \bibinfo {pages} {L201102} (\bibinfo {year} {2024})}\BibitemShut {NoStop}%
\bibitem [{\citenamefont
  {Chen}(2023{\natexlab{a}})}]{weichen_spincurvature_2023}%
  \BibitemOpen
  \bibfield  {author} {\bibinfo {author} {\bibfnamefont {W.}~\bibnamefont
  {Chen}},\ }\bibfield  {title} {\bibinfo {title} {Optical absorption
  measurement of spin {B}erry curvature and spin {C}hern marker},\ }\href
  {https://doi.org/10.1088/1361-648X/acba72} {\bibfield  {journal} {\bibinfo
  {journal} {Journal of Physics: Condensed Matter}\ }\textbf {\bibinfo {volume}
  {35}},\ \bibinfo {pages} {155601} (\bibinfo {year}
  {2023}{\natexlab{a}})}\BibitemShut {NoStop}%
\bibitem [{\citenamefont {Li}\ and\ \citenamefont {Mong}(2019)}]{li_prb_2019}%
  \BibitemOpen
  \bibfield  {author} {\bibinfo {author} {\bibfnamefont {Z.}~\bibnamefont
  {Li}}\ and\ \bibinfo {author} {\bibfnamefont {R.~S.~K.}\ \bibnamefont
  {Mong}},\ }\bibfield  {title} {\bibinfo {title} {Local formula for the
  $\mathbb{Z}_{2}$ invariant of topological insulators},\ }\href
  {https://doi.org/10.1103/PhysRevB.100.205101} {\bibfield  {journal} {\bibinfo
   {journal} {Phys. Rev. B}\ }\textbf {\bibinfo {volume} {100}},\ \bibinfo
  {pages} {205101} (\bibinfo {year} {2019})}\BibitemShut {NoStop}%
\bibitem [{\citenamefont {Cerjan}\ \emph {et~al.}(2024)\citenamefont {Cerjan},
  \citenamefont {Loring},\ and\ \citenamefont {Schulz-Baldes}}]{lcti_prl_2024}%
  \BibitemOpen
  \bibfield  {author} {\bibinfo {author} {\bibfnamefont {A.}~\bibnamefont
  {Cerjan}}, \bibinfo {author} {\bibfnamefont {T.~A.}\ \bibnamefont {Loring}},\
  and\ \bibinfo {author} {\bibfnamefont {H.}~\bibnamefont {Schulz-Baldes}},\
  }\bibfield  {title} {\bibinfo {title} {{Local Markers for Crystalline
  Topology}},\ }\href {https://doi.org/10.1103/PhysRevLett.132.073803}
  {\bibfield  {journal} {\bibinfo  {journal} {Phys. Rev. Lett.}\ }\textbf
  {\bibinfo {volume} {132}},\ \bibinfo {pages} {073803} (\bibinfo {year}
  {2024})}\BibitemShut {NoStop}%
\bibitem [{\citenamefont {Lin}\ \emph {et~al.}(2024)\citenamefont {Lin},
  \citenamefont {Palumbo}, \citenamefont {Guo}, \citenamefont {Hwang},
  \citenamefont {Blackburn}, \citenamefont {Shoemaker}, \citenamefont
  {Mahmood}, \citenamefont {Wang}, \citenamefont {Fiete}, \citenamefont
  {Wieder},\ and\ \citenamefont {Bradlyn}}]{layerlscm_natcomm_2024}%
  \BibitemOpen
  \bibfield  {author} {\bibinfo {author} {\bibfnamefont {K.-S.}\ \bibnamefont
  {Lin}}, \bibinfo {author} {\bibfnamefont {G.}~\bibnamefont {Palumbo}},
  \bibinfo {author} {\bibfnamefont {Z.}~\bibnamefont {Guo}}, \bibinfo {author}
  {\bibfnamefont {Y.}~\bibnamefont {Hwang}}, \bibinfo {author} {\bibfnamefont
  {J.}~\bibnamefont {Blackburn}}, \bibinfo {author} {\bibfnamefont {D.~P.}\
  \bibnamefont {Shoemaker}}, \bibinfo {author} {\bibfnamefont {F.}~\bibnamefont
  {Mahmood}}, \bibinfo {author} {\bibfnamefont {Z.}~\bibnamefont {Wang}},
  \bibinfo {author} {\bibfnamefont {G.~A.}\ \bibnamefont {Fiete}}, \bibinfo
  {author} {\bibfnamefont {B.~J.}\ \bibnamefont {Wieder}},\ and\ \bibinfo
  {author} {\bibfnamefont {B.}~\bibnamefont {Bradlyn}},\ }\bibfield  {title}
  {\bibinfo {title} {Spin-resolved topology and partial axion angles in
  three-dimensional insulators},\ }\href
  {https://doi.org/10.1038/s41467-024-44762-w} {\bibfield  {journal} {\bibinfo
  {journal} {Nature Communications}\ }\textbf {\bibinfo {volume} {15}},\
  \bibinfo {pages} {550} (\bibinfo {year} {2024})}\BibitemShut {NoStop}%
\bibitem [{\citenamefont {Kim}\ \emph {et~al.}(2023)\citenamefont {Kim},
  \citenamefont {Jeon}, \citenamefont {Park},\ and\ \citenamefont
  {Kim}}]{higherordermarker_npj_2023}%
  \BibitemOpen
  \bibfield  {author} {\bibinfo {author} {\bibfnamefont {S.-W.}\ \bibnamefont
  {Kim}}, \bibinfo {author} {\bibfnamefont {S.}~\bibnamefont {Jeon}}, \bibinfo
  {author} {\bibfnamefont {M.~J.}\ \bibnamefont {Park}},\ and\ \bibinfo
  {author} {\bibfnamefont {Y.}~\bibnamefont {Kim}},\ }\bibfield  {title}
  {\bibinfo {title} {{Replica higher-order topology of Hofstadter butterflies
  in twisted bilayer graphene}},\ }\href
  {https://doi.org/10.1038/s41524-023-01105-5} {\bibfield  {journal} {\bibinfo
  {journal} {npj Computational Materials}\ }\textbf {\bibinfo {volume} {9}},\
  \bibinfo {pages} {152} (\bibinfo {year} {2023})}\BibitemShut {NoStop}%
\bibitem [{\citenamefont {Ba\`u}\ and\ \citenamefont
  {Marrazzo}(2024)}]{pbclcm}%
  \BibitemOpen
  \bibfield  {author} {\bibinfo {author} {\bibfnamefont {N.}~\bibnamefont
  {Ba\`u}}\ and\ \bibinfo {author} {\bibfnamefont {A.}~\bibnamefont
  {Marrazzo}},\ }\bibfield  {title} {\bibinfo {title} {Local {C}hern marker for
  periodic systems},\ }\href {https://doi.org/10.1103/PhysRevB.109.014206}
  {\bibfield  {journal} {\bibinfo  {journal} {Phys. Rev. B}\ }\textbf {\bibinfo
  {volume} {109}},\ \bibinfo {pages} {014206} (\bibinfo {year}
  {2024})}\BibitemShut {NoStop}%
\bibitem [{\citenamefont {Pozo}\ \emph {et~al.}(2019)\citenamefont {Pozo},
  \citenamefont {Repellin},\ and\ \citenamefont {Grushin}}]{Pozo_PRL_2019}%
  \BibitemOpen
  \bibfield  {author} {\bibinfo {author} {\bibfnamefont {O.}~\bibnamefont
  {Pozo}}, \bibinfo {author} {\bibfnamefont {C.}~\bibnamefont {Repellin}},\
  and\ \bibinfo {author} {\bibfnamefont {A.~G.}\ \bibnamefont {Grushin}},\
  }\bibfield  {title} {\bibinfo {title} {{Quantization in Chiral Higher Order
  Topological Insulators: Circular Dichroism and Local Chern Marker}},\ }\href
  {https://doi.org/10.1103/PhysRevLett.123.247401} {\bibfield  {journal}
  {\bibinfo  {journal} {Phys. Rev. Lett.}\ }\textbf {\bibinfo {volume} {123}},\
  \bibinfo {pages} {247401} (\bibinfo {year} {2019})}\BibitemShut {NoStop}%
\bibitem [{\citenamefont {Marsal}\ \emph
  {et~al.}(2020{\natexlab{b}})\citenamefont {Marsal}, \citenamefont {Varjas},\
  and\ \citenamefont {Grushin}}]{Marsal_PNAS_2020}%
  \BibitemOpen
  \bibfield  {author} {\bibinfo {author} {\bibfnamefont {Q.}~\bibnamefont
  {Marsal}}, \bibinfo {author} {\bibfnamefont {D.}~\bibnamefont {Varjas}},\
  and\ \bibinfo {author} {\bibfnamefont {A.~G.}\ \bibnamefont {Grushin}},\
  }\bibfield  {title} {\bibinfo {title} {{Topological Weaire-Thorpe models of
  amorphous matter}},\ }\href {https://doi.org/10.1073/pnas.2007384117}
  {\bibfield  {journal} {\bibinfo  {journal} {Proceedings of the National
  Academy of Sciences}\ }\textbf {\bibinfo {volume} {117}},\ \bibinfo {pages}
  {30260} (\bibinfo {year} {2020}{\natexlab{b}})}\BibitemShut {NoStop}%
\bibitem [{\citenamefont {Molignini}\ \emph {et~al.}(2023)\citenamefont
  {Molignini}, \citenamefont {Lapierre}, \citenamefont {Chitra},\ and\
  \citenamefont {Chen}}]{Molignini_scipost_2023}%
  \BibitemOpen
  \bibfield  {author} {\bibinfo {author} {\bibfnamefont {P.}~\bibnamefont
  {Molignini}}, \bibinfo {author} {\bibfnamefont {B.}~\bibnamefont {Lapierre}},
  \bibinfo {author} {\bibfnamefont {R.}~\bibnamefont {Chitra}},\ and\ \bibinfo
  {author} {\bibfnamefont {W.}~\bibnamefont {Chen}},\ }\bibfield  {title}
  {\bibinfo {title} {{Probing Chern number by opacity and topological phase
  transition by a nonlocal Chern marker}},\ }\href
  {https://doi.org/10.21468/SciPostPhysCore.6.3.059} {\bibfield  {journal}
  {\bibinfo  {journal} {SciPost Phys. Core}\ }\textbf {\bibinfo {volume} {6}},\
  \bibinfo {pages} {059} (\bibinfo {year} {2023})}\BibitemShut {NoStop}%
\bibitem [{\citenamefont {Hannukainen}\ \emph {et~al.}(2022)\citenamefont
  {Hannukainen}, \citenamefont {Mart\'{\i}nez}, \citenamefont {Bardarson},\
  and\ \citenamefont {Kvorning}}]{hannukainen_2022}%
  \BibitemOpen
  \bibfield  {author} {\bibinfo {author} {\bibfnamefont {J.~D.}\ \bibnamefont
  {Hannukainen}}, \bibinfo {author} {\bibfnamefont {M.~F.}\ \bibnamefont
  {Mart\'{\i}nez}}, \bibinfo {author} {\bibfnamefont {J.~H.}\ \bibnamefont
  {Bardarson}},\ and\ \bibinfo {author} {\bibfnamefont {T.~K.}\ \bibnamefont
  {Kvorning}},\ }\bibfield  {title} {\bibinfo {title} {Local {T}opological
  {M}arkers in {O}dd {S}patial {D}imensions and {T}heir {A}pplication to
  {A}morphous {T}opological {M}atter},\ }\href
  {https://doi.org/10.1103/PhysRevLett.129.277601} {\bibfield  {journal}
  {\bibinfo  {journal} {Phys. Rev. Lett.}\ }\textbf {\bibinfo {volume} {129}},\
  \bibinfo {pages} {277601} (\bibinfo {year} {2022})}\BibitemShut {NoStop}%
\bibitem [{\citenamefont {Ryu}\ \emph {et~al.}(2010)\citenamefont {Ryu},
  \citenamefont {Schnyder}, \citenamefont {Furusaki},\ and\ \citenamefont
  {Ludwig}}]{tenfoldway_2010}%
  \BibitemOpen
  \bibfield  {author} {\bibinfo {author} {\bibfnamefont {S.}~\bibnamefont
  {Ryu}}, \bibinfo {author} {\bibfnamefont {A.~P.}\ \bibnamefont {Schnyder}},
  \bibinfo {author} {\bibfnamefont {A.}~\bibnamefont {Furusaki}},\ and\
  \bibinfo {author} {\bibfnamefont {A.~W.~W.}\ \bibnamefont {Ludwig}},\
  }\bibfield  {title} {\bibinfo {title} {Topological insulators and
  superconductors: tenfold way and dimensional hierarchy},\ }\href
  {https://doi.org/10.1088/1367-2630/12/6/065010} {\bibfield  {journal}
  {\bibinfo  {journal} {New Journal of Physics}\ }\textbf {\bibinfo {volume}
  {12}},\ \bibinfo {pages} {065010} (\bibinfo {year} {2010})}\BibitemShut
  {NoStop}%
\bibitem [{\citenamefont {Chen}(2023{\natexlab{b}})}]{wei_prb_2022}%
  \BibitemOpen
  \bibfield  {author} {\bibinfo {author} {\bibfnamefont {W.}~\bibnamefont
  {Chen}},\ }\bibfield  {title} {\bibinfo {title} {Universal topological
  marker},\ }\href {https://doi.org/10.1103/PhysRevB.107.045111} {\bibfield
  {journal} {\bibinfo  {journal} {Phys. Rev. B}\ }\textbf {\bibinfo {volume}
  {107}},\ \bibinfo {pages} {045111} (\bibinfo {year}
  {2023}{\natexlab{b}})}\BibitemShut {NoStop}%
\bibitem [{\citenamefont {Prodan}(2009)}]{prodan_2009}%
  \BibitemOpen
  \bibfield  {author} {\bibinfo {author} {\bibfnamefont {E.}~\bibnamefont
  {Prodan}},\ }\bibfield  {title} {\bibinfo {title} {Robustness of the
  spin-{C}hern number},\ }\href {https://doi.org/10.1103/PhysRevB.80.125327}
  {\bibfield  {journal} {\bibinfo  {journal} {Phys. Rev. B}\ }\textbf {\bibinfo
  {volume} {80}},\ \bibinfo {pages} {125327} (\bibinfo {year}
  {2009})}\BibitemShut {NoStop}%
\bibitem [{\citenamefont {Soluyanov}\ and\ \citenamefont
  {Vanderbilt}(2011{\natexlab{b}})}]{soluyanov_prb_2011}%
  \BibitemOpen
  \bibfield  {author} {\bibinfo {author} {\bibfnamefont {A.~A.}\ \bibnamefont
  {Soluyanov}}\ and\ \bibinfo {author} {\bibfnamefont {D.}~\bibnamefont
  {Vanderbilt}},\ }\bibfield  {title} {\bibinfo {title} {Computing topological
  invariants without inversion symmetry},\ }\href
  {https://doi.org/10.1103/PhysRevB.83.235401} {\bibfield  {journal} {\bibinfo
  {journal} {Phys. Rev. B}\ }\textbf {\bibinfo {volume} {83}},\ \bibinfo
  {pages} {235401} (\bibinfo {year} {2011}{\natexlab{b}})}\BibitemShut
  {NoStop}%
\bibitem [{\citenamefont {Soluyanov}\ and\ \citenamefont
  {Vanderbilt}(2012)}]{soluyanov_2012}%
  \BibitemOpen
  \bibfield  {author} {\bibinfo {author} {\bibfnamefont {A.~A.}\ \bibnamefont
  {Soluyanov}}\ and\ \bibinfo {author} {\bibfnamefont {D.}~\bibnamefont
  {Vanderbilt}},\ }\bibfield  {title} {\bibinfo {title} {Smooth gauge for
  topological insulators},\ }\href {https://doi.org/10.1103/PhysRevB.85.115415}
  {\bibfield  {journal} {\bibinfo  {journal} {Phys. Rev. B}\ }\textbf {\bibinfo
  {volume} {85}},\ \bibinfo {pages} {115415} (\bibinfo {year}
  {2012})}\BibitemShut {NoStop}%
\bibitem [{\citenamefont {Bellissard}\ \emph {et~al.}(1994)\citenamefont
  {Bellissard}, \citenamefont {van Elst},\ and\ \citenamefont
  {Schulz-Baldes}}]{bellissard_noncommutative}%
  \BibitemOpen
  \bibfield  {author} {\bibinfo {author} {\bibfnamefont {J.}~\bibnamefont
  {Bellissard}}, \bibinfo {author} {\bibfnamefont {A.}~\bibnamefont {van
  Elst}},\ and\ \bibinfo {author} {\bibfnamefont {H.}~\bibnamefont
  {Schulz-Baldes}},\ }\bibfield  {title} {\bibinfo {title} {{The noncommutative
  geometry of the quantum {H}all effect}},\ }\href
  {https://doi.org/10.1063/1.530758} {\bibfield  {journal} {\bibinfo  {journal}
  {Journal of Mathematical Physics}\ }\textbf {\bibinfo {volume} {35}},\
  \bibinfo {pages} {5373} (\bibinfo {year} {1994})}\BibitemShut {NoStop}%
\bibitem [{\citenamefont {Prodan}(2010)}]{prodan_2010}%
  \BibitemOpen
  \bibfield  {author} {\bibinfo {author} {\bibfnamefont {E.}~\bibnamefont
  {Prodan}},\ }\bibfield  {title} {\bibinfo {title} {Non-commutative tools for
  topological insulators},\ }\href
  {https://doi.org/10.1088/1367-2630/12/6/065003} {\bibfield  {journal}
  {\bibinfo  {journal} {New Journal of Physics}\ }\textbf {\bibinfo {volume}
  {12}},\ \bibinfo {pages} {065003} (\bibinfo {year} {2010})}\BibitemShut
  {NoStop}%
\bibitem [{\citenamefont {Soluyanov}(2012)}]{soluyanov_phd}%
  \BibitemOpen
  \bibfield  {author} {\bibinfo {author} {\bibfnamefont {A.~A.}\ \bibnamefont
  {Soluyanov}},\ }\bibfield  {title} {\bibinfo {title} {Topological aspects of
  band theory},\ }\bibfield  {journal} {\bibinfo  {journal} {PhD thesis,
  Rutgers University}\ }\href {https://doi.org/10.7282/T3N0159G}
  {10.7282/T3N0159G} (\bibinfo {year} {2012})\BibitemShut {NoStop}%
\bibitem [{\citenamefont {Marrazzo}\ and\ \citenamefont
  {Resta}(2017)}]{marrazzo_prb_2017}%
  \BibitemOpen
  \bibfield  {author} {\bibinfo {author} {\bibfnamefont {A.}~\bibnamefont
  {Marrazzo}}\ and\ \bibinfo {author} {\bibfnamefont {R.}~\bibnamefont
  {Resta}},\ }\bibfield  {title} {\bibinfo {title} {Locality of the anomalous
  {H}all conductivity},\ }\href {https://doi.org/10.1103/PhysRevB.95.121114}
  {\bibfield  {journal} {\bibinfo  {journal} {Phys. Rev. B}\ }\textbf {\bibinfo
  {volume} {95}},\ \bibinfo {pages} {121114(R)} (\bibinfo {year}
  {2017})}\BibitemShut {NoStop}%
\bibitem [{\citenamefont {Rauch}\ \emph {et~al.}(2018)\citenamefont {Rauch},
  \citenamefont {Olsen}, \citenamefont {Vanderbilt},\ and\ \citenamefont
  {Souza}}]{rauch_ahc_pbc2018}%
  \BibitemOpen
  \bibfield  {author} {\bibinfo {author} {\bibfnamefont {T.}~\bibnamefont
  {Rauch}}, \bibinfo {author} {\bibfnamefont {T.}~\bibnamefont {Olsen}},
  \bibinfo {author} {\bibfnamefont {D.}~\bibnamefont {Vanderbilt}},\ and\
  \bibinfo {author} {\bibfnamefont {I.}~\bibnamefont {Souza}},\ }\bibfield
  {title} {\bibinfo {title} {Geometric and nongeometric contributions to the
  surface anomalous {H}all conductivity},\ }\href
  {https://doi.org/10.1103/PhysRevB.98.115108} {\bibfield  {journal} {\bibinfo
  {journal} {Phys. Rev. B}\ }\textbf {\bibinfo {volume} {98}},\ \bibinfo
  {pages} {115108} (\bibinfo {year} {2018})}\BibitemShut {NoStop}%
\bibitem [{\citenamefont {Resta}(1998)}]{resta_prl_1998}%
  \BibitemOpen
  \bibfield  {author} {\bibinfo {author} {\bibfnamefont {R.}~\bibnamefont
  {Resta}},\ }\bibfield  {title} {\bibinfo {title} {Quantum-{M}echanical
  {P}osition {O}perator in {E}xtended {S}ystems},\ }\href
  {https://doi.org/10.1103/PhysRevLett.80.1800} {\bibfield  {journal} {\bibinfo
   {journal} {Phys. Rev. Lett.}\ }\textbf {\bibinfo {volume} {80}},\ \bibinfo
  {pages} {1800} (\bibinfo {year} {1998})}\BibitemShut {NoStop}%
\bibitem [{\citenamefont {Souza}\ \emph {et~al.}(2004)\citenamefont {Souza},
  \citenamefont {\'I\~niguez},\ and\ \citenamefont
  {Vanderbilt}}]{souza_covariantderivative_prb_2004}%
  \BibitemOpen
  \bibfield  {author} {\bibinfo {author} {\bibfnamefont {I.}~\bibnamefont
  {Souza}}, \bibinfo {author} {\bibfnamefont {J.}~\bibnamefont {\'I\~niguez}},\
  and\ \bibinfo {author} {\bibfnamefont {D.}~\bibnamefont {Vanderbilt}},\
  }\bibfield  {title} {\bibinfo {title} {Dynamics of {B}erry-phase polarization
  in time-dependent electric fields},\ }\href
  {https://doi.org/10.1103/PhysRevB.69.085106} {\bibfield  {journal} {\bibinfo
  {journal} {Phys. Rev. B}\ }\textbf {\bibinfo {volume} {69}},\ \bibinfo
  {pages} {085106} (\bibinfo {year} {2004})}\BibitemShut {NoStop}%
\bibitem [{\citenamefont {Teo}\ \emph {et~al.}(2008)\citenamefont {Teo},
  \citenamefont {Fu},\ and\ \citenamefont {Kane}}]{mirrorcherndef_prb_2008}%
  \BibitemOpen
  \bibfield  {author} {\bibinfo {author} {\bibfnamefont {J.~C.~Y.}\
  \bibnamefont {Teo}}, \bibinfo {author} {\bibfnamefont {L.}~\bibnamefont
  {Fu}},\ and\ \bibinfo {author} {\bibfnamefont {C.~L.}\ \bibnamefont {Kane}},\
  }\bibfield  {title} {\bibinfo {title} {Surface states and topological
  invariants in three-dimensional topological insulators: Application to
  $\text{Bi}_{1\ensuremath{-}x}\text{Sb}_{x}$},\ }\href
  {https://doi.org/10.1103/PhysRevB.78.045426} {\bibfield  {journal} {\bibinfo
  {journal} {Phys. Rev. B}\ }\textbf {\bibinfo {volume} {78}},\ \bibinfo
  {pages} {045426} (\bibinfo {year} {2008})}\BibitemShut {NoStop}%
\bibitem [{\citenamefont {Rauch}\ \emph {et~al.}(2020)\citenamefont {Rauch},
  \citenamefont {T\"opler},\ and\ \citenamefont {Mertig}}]{rauch_prb_2020}%
  \BibitemOpen
  \bibfield  {author} {\bibinfo {author} {\bibfnamefont {T.}~\bibnamefont
  {Rauch}}, \bibinfo {author} {\bibfnamefont {F.}~\bibnamefont {T\"opler}},\
  and\ \bibinfo {author} {\bibfnamefont {I.}~\bibnamefont {Mertig}},\
  }\bibfield  {title} {\bibinfo {title} {Local spin {H}all conductivity},\
  }\href {https://doi.org/10.1103/PhysRevB.101.064206} {\bibfield  {journal}
  {\bibinfo  {journal} {Phys. Rev. B}\ }\textbf {\bibinfo {volume} {101}},\
  \bibinfo {pages} {064206} (\bibinfo {year} {2020})}\BibitemShut {NoStop}%
\bibitem [{\citenamefont {Rauch}\ \emph {et~al.}(2021)\citenamefont {Rauch},
  \citenamefont {Olsen}, \citenamefont {Vanderbilt},\ and\ \citenamefont
  {Souza}}]{mirrorchern_prb_2021}%
  \BibitemOpen
  \bibfield  {author} {\bibinfo {author} {\bibfnamefont {T.}~\bibnamefont
  {Rauch}}, \bibinfo {author} {\bibfnamefont {T.}~\bibnamefont {Olsen}},
  \bibinfo {author} {\bibfnamefont {D.}~\bibnamefont {Vanderbilt}},\ and\
  \bibinfo {author} {\bibfnamefont {I.}~\bibnamefont {Souza}},\ }\bibfield
  {title} {\bibinfo {title} {Mirror {C}hern numbers in the hybrid {W}annier
  representation},\ }\href {https://doi.org/10.1103/PhysRevB.103.195103}
  {\bibfield  {journal} {\bibinfo  {journal} {Phys. Rev. B}\ }\textbf {\bibinfo
  {volume} {103}},\ \bibinfo {pages} {195103} (\bibinfo {year}
  {2021})}\BibitemShut {NoStop}%
\bibitem [{\citenamefont {Marzari}\ and\ \citenamefont
  {Vanderbilt}(1997)}]{marzari_prb_1997}%
  \BibitemOpen
  \bibfield  {author} {\bibinfo {author} {\bibfnamefont {N.}~\bibnamefont
  {Marzari}}\ and\ \bibinfo {author} {\bibfnamefont {D.}~\bibnamefont
  {Vanderbilt}},\ }\bibfield  {title} {\bibinfo {title} {Maximally localized
  generalized {W}annier functions for composite energy bands},\ }\href
  {https://doi.org/10.1103/PhysRevB.56.12847} {\bibfield  {journal} {\bibinfo
  {journal} {Phys. Rev. B}\ }\textbf {\bibinfo {volume} {56}},\ \bibinfo
  {pages} {12847} (\bibinfo {year} {1997})}\BibitemShut {NoStop}%
\bibitem [{\citenamefont {Marzari}\ \emph {et~al.}(2012)\citenamefont
  {Marzari}, \citenamefont {Mostofi}, \citenamefont {Yates}, \citenamefont
  {Souza},\ and\ \citenamefont {Vanderbilt}}]{marzari_wf_review}%
  \BibitemOpen
  \bibfield  {author} {\bibinfo {author} {\bibfnamefont {N.}~\bibnamefont
  {Marzari}}, \bibinfo {author} {\bibfnamefont {A.~A.}\ \bibnamefont
  {Mostofi}}, \bibinfo {author} {\bibfnamefont {J.~R.}\ \bibnamefont {Yates}},
  \bibinfo {author} {\bibfnamefont {I.}~\bibnamefont {Souza}},\ and\ \bibinfo
  {author} {\bibfnamefont {D.}~\bibnamefont {Vanderbilt}},\ }\bibfield  {title}
  {\bibinfo {title} {Maximally localized {W}annier functions: {T}heory and
  applications},\ }\href {https://doi.org/10.1103/RevModPhys.84.1419}
  {\bibfield  {journal} {\bibinfo  {journal} {Rev. Mod. Phys.}\ }\textbf
  {\bibinfo {volume} {84}},\ \bibinfo {pages} {1419} (\bibinfo {year}
  {2012})}\BibitemShut {NoStop}%
\bibitem [{\citenamefont {Boys}(1960)}]{boys_obcwf}%
  \BibitemOpen
  \bibfield  {author} {\bibinfo {author} {\bibfnamefont {S.~F.}\ \bibnamefont
  {Boys}},\ }\bibfield  {title} {\bibinfo {title} {Construction of {S}ome
  {M}olecular {O}rbitals to {B}e {A}pproximately {I}nvariant for {C}hanges from
  {O}ne {M}olecule to {A}nother},\ }\href
  {https://doi.org/10.1103/RevModPhys.32.296} {\bibfield  {journal} {\bibinfo
  {journal} {Rev. Mod. Phys.}\ }\textbf {\bibinfo {volume} {32}},\ \bibinfo
  {pages} {296} (\bibinfo {year} {1960})}\BibitemShut {NoStop}%
\bibitem [{\citenamefont {Winkler}\ \emph {et~al.}(2016)\citenamefont
  {Winkler}, \citenamefont {Soluyanov},\ and\ \citenamefont
  {Troyer}}]{wannier_z2_dimensions_soluyanov}%
  \BibitemOpen
  \bibfield  {author} {\bibinfo {author} {\bibfnamefont {G.~W.}\ \bibnamefont
  {Winkler}}, \bibinfo {author} {\bibfnamefont {A.~A.}\ \bibnamefont
  {Soluyanov}},\ and\ \bibinfo {author} {\bibfnamefont {M.}~\bibnamefont
  {Troyer}},\ }\bibfield  {title} {\bibinfo {title} {{Smooth gauge and Wannier
  functions for topological band structures in arbitrary dimensions}},\ }\href
  {https://doi.org/10.1103/PhysRevB.93.035453} {\bibfield  {journal} {\bibinfo
  {journal} {Phys. Rev. B}\ }\textbf {\bibinfo {volume} {93}},\ \bibinfo
  {pages} {035453} (\bibinfo {year} {2016})}\BibitemShut {NoStop}%
\bibitem [{Note1()}]{Note1}%
  \BibitemOpen
  \bibinfo {note} {\label {footnote:1} The spillage can be interpreted as a
  measure of how much two projectors describe the same subspace, or
  equivalently, a distance between manifolds. Since the trace of $\protect
  \mathcal P$ is equal to the number of occupied states $N_{occ}$, this
  formulation of the spillage is equivalent to the more common expression in
  terms of the projector onto empty states: $\protect \mathrm {Tr}[(\protect
  \mathcal P-\protect \mathcal P_{\Theta })^2]/(2N_{occ})=\protect \mathrm
  {Tr}[\protect \mathcal P\protect \mathcal Q_{\Theta }]/N_{occ}=\protect
  \mathrm {Tr}[\protect \mathcal Q\protect \mathcal P_{\Theta }]/N_{occ}$. From
  these forms it is easier to see that, for $\gamma =0$, the projectors
  describe the same subspace, while if $\gamma =1$, the two projectors are
  orthogonal and thus describe orthogonal subspaces.}\BibitemShut {Stop}%
\bibitem [{\citenamefont {Brouder}\ \emph {et~al.}(2007)\citenamefont
  {Brouder}, \citenamefont {Panati}, \citenamefont {Calandra}, \citenamefont
  {Mourougane},\ and\ \citenamefont {Marzari}}]{marzari_prl_2007}%
  \BibitemOpen
  \bibfield  {author} {\bibinfo {author} {\bibfnamefont {C.}~\bibnamefont
  {Brouder}}, \bibinfo {author} {\bibfnamefont {G.}~\bibnamefont {Panati}},
  \bibinfo {author} {\bibfnamefont {M.}~\bibnamefont {Calandra}}, \bibinfo
  {author} {\bibfnamefont {C.}~\bibnamefont {Mourougane}},\ and\ \bibinfo
  {author} {\bibfnamefont {N.}~\bibnamefont {Marzari}},\ }\bibfield  {title}
  {\bibinfo {title} {Exponential {L}ocalization of {W}annier {F}unctions in
  {I}nsulators},\ }\href {https://doi.org/10.1103/PhysRevLett.98.046402}
  {\bibfield  {journal} {\bibinfo  {journal} {Phys. Rev. Lett.}\ }\textbf
  {\bibinfo {volume} {98}},\ \bibinfo {pages} {046402} (\bibinfo {year}
  {2007})}\BibitemShut {NoStop}%
\bibitem [{\citenamefont {Souza}\ \emph {et~al.}(2001)\citenamefont {Souza},
  \citenamefont {Marzari},\ and\ \citenamefont
  {Vanderbilt}}]{souza_disentanglement_prb_2001}%
  \BibitemOpen
  \bibfield  {author} {\bibinfo {author} {\bibfnamefont {I.}~\bibnamefont
  {Souza}}, \bibinfo {author} {\bibfnamefont {N.}~\bibnamefont {Marzari}},\
  and\ \bibinfo {author} {\bibfnamefont {D.}~\bibnamefont {Vanderbilt}},\
  }\bibfield  {title} {\bibinfo {title} {Maximally localized {W}annier
  functions for entangled energy bands},\ }\href
  {https://doi.org/10.1103/PhysRevB.65.035109} {\bibfield  {journal} {\bibinfo
  {journal} {Phys. Rev. B}\ }\textbf {\bibinfo {volume} {65}},\ \bibinfo
  {pages} {035109} (\bibinfo {year} {2001})}\BibitemShut {NoStop}%
\bibitem [{\citenamefont {Anderson}(1958)}]{anderson_disorder_1958}%
  \BibitemOpen
  \bibfield  {author} {\bibinfo {author} {\bibfnamefont {P.~W.}\ \bibnamefont
  {Anderson}},\ }\bibfield  {title} {\bibinfo {title} {{Absence of Diffusion in
  Certain Random Lattices}},\ }\href {https://doi.org/10.1103/PhysRev.109.1492}
  {\bibfield  {journal} {\bibinfo  {journal} {Phys. Rev.}\ }\textbf {\bibinfo
  {volume} {109}},\ \bibinfo {pages} {1492} (\bibinfo {year}
  {1958})}\BibitemShut {NoStop}%
\bibitem [{\citenamefont {Yang}\ \emph {et~al.}(2011)\citenamefont {Yang},
  \citenamefont {Xu}, \citenamefont {Sheng}, \citenamefont {Wang},
  \citenamefont {Xing},\ and\ \citenamefont {Sheng}}]{yang_prl_2011}%
  \BibitemOpen
  \bibfield  {author} {\bibinfo {author} {\bibfnamefont {Y.}~\bibnamefont
  {Yang}}, \bibinfo {author} {\bibfnamefont {Z.}~\bibnamefont {Xu}}, \bibinfo
  {author} {\bibfnamefont {L.}~\bibnamefont {Sheng}}, \bibinfo {author}
  {\bibfnamefont {B.}~\bibnamefont {Wang}}, \bibinfo {author} {\bibfnamefont
  {D.~Y.}\ \bibnamefont {Xing}},\ and\ \bibinfo {author} {\bibfnamefont
  {D.~N.}\ \bibnamefont {Sheng}},\ }\bibfield  {title} {\bibinfo {title}
  {{Time-Reversal-Symmetry-Broken Quantum Spin Hall Effect}},\ }\href
  {https://doi.org/10.1103/PhysRevLett.107.066602} {\bibfield  {journal}
  {\bibinfo  {journal} {Phys. Rev. Lett.}\ }\textbf {\bibinfo {volume} {107}},\
  \bibinfo {pages} {066602} (\bibinfo {year} {2011})}\BibitemShut {NoStop}%
\bibitem [{str()}]{strawberrypy}%
  \BibitemOpen
  \href@noop {} {}\bibinfo {note} {The \textsc{StraWBerryPy} code package is
  available at
  \url{https://strawberrypy.readthedocs.io/en/latest/}}\BibitemShut {NoStop}%
\bibitem [{\citenamefont {Marrazzo}\ \emph {et~al.}(2023)\citenamefont
  {Marrazzo}, \citenamefont {Beck}, \citenamefont {Margine}, \citenamefont
  {Marzari}, \citenamefont {Mostofi}, \citenamefont {Qiao}, \citenamefont
  {Souza}, \citenamefont {Tsirkin}, \citenamefont {Yates},\ and\ \citenamefont
  {Pizzi}}]{WF_ecosys_2023}%
  \BibitemOpen
  \bibfield  {author} {\bibinfo {author} {\bibfnamefont {A.}~\bibnamefont
  {Marrazzo}}, \bibinfo {author} {\bibfnamefont {S.}~\bibnamefont {Beck}},
  \bibinfo {author} {\bibfnamefont {E.~R.}\ \bibnamefont {Margine}}, \bibinfo
  {author} {\bibfnamefont {N.}~\bibnamefont {Marzari}}, \bibinfo {author}
  {\bibfnamefont {A.~A.}\ \bibnamefont {Mostofi}}, \bibinfo {author}
  {\bibfnamefont {J.}~\bibnamefont {Qiao}}, \bibinfo {author} {\bibfnamefont
  {I.}~\bibnamefont {Souza}}, \bibinfo {author} {\bibfnamefont {S.~S.}\
  \bibnamefont {Tsirkin}}, \bibinfo {author} {\bibfnamefont {J.~R.}\
  \bibnamefont {Yates}},\ and\ \bibinfo {author} {\bibfnamefont
  {G.}~\bibnamefont {Pizzi}},\ }\href@noop {} {\bibinfo {title} {{The
  Wannier-Functions Software Ecosystem for Materials Simulations}}} (\bibinfo
  {year} {2023}),\ \Eprint {https://arxiv.org/abs/2312.10769} {arXiv:2312.10769
  [cond-mat.mtrl-sci]} \BibitemShut {NoStop}%
\bibitem [{tbm()}]{tbmodels}%
  \BibitemOpen
  \href@noop {} {}\bibinfo {note} {The \textsc{TBmodels} code package is
  available at
  \url{https://tbmodels.greschd.ch/en/latest/index.html}}\BibitemShut {NoStop}%
\bibitem [{\citenamefont {Gresch}\ \emph {et~al.}(2018)\citenamefont {Gresch},
  \citenamefont {Wu}, \citenamefont {Winkler}, \citenamefont {H\"auselmann},
  \citenamefont {Troyer},\ and\ \citenamefont {Soluyanov}}]{TB}%
  \BibitemOpen
  \bibfield  {author} {\bibinfo {author} {\bibfnamefont {D.}~\bibnamefont
  {Gresch}}, \bibinfo {author} {\bibfnamefont {Q.~S.}\ \bibnamefont {Wu}},
  \bibinfo {author} {\bibfnamefont {G.~W.}\ \bibnamefont {Winkler}}, \bibinfo
  {author} {\bibfnamefont {R.}~\bibnamefont {H\"auselmann}}, \bibinfo {author}
  {\bibfnamefont {M.}~\bibnamefont {Troyer}},\ and\ \bibinfo {author}
  {\bibfnamefont {A.~A.}\ \bibnamefont {Soluyanov}},\ }\bibfield  {title}
  {\bibinfo {title} {Automated construction of symmetrized {W}annier-like
  tight-binding models from ab initio calculations},\ }\href
  {https://doi.org/10.1103/PhysRevMaterials.2.103805} {\bibfield  {journal}
  {\bibinfo  {journal} {Phys. Rev. Mater.}\ }\textbf {\bibinfo {volume} {2}},\
  \bibinfo {pages} {103805} (\bibinfo {year} {2018})}\BibitemShut {NoStop}%
\bibitem [{pyt()}]{pythtb}%
  \BibitemOpen
  \href@noop {} {}\bibinfo {note} {The \textsc{PythTB} code package is
  available at
  \url{http://www.physics.rutgers.edu/pythtb/about.html}}\BibitemShut {NoStop}%
\bibitem [{\citenamefont {Corbae}\ \emph
  {et~al.}(2023{\natexlab{a}})\citenamefont {Corbae}, \citenamefont
  {Hannukainen}, \citenamefont {Marsal}, \citenamefont {noz Segovia},\ and\
  \citenamefont {Grushin}}]{corbae_EPL_2023}%
  \BibitemOpen
  \bibfield  {author} {\bibinfo {author} {\bibfnamefont {P.}~\bibnamefont
  {Corbae}}, \bibinfo {author} {\bibfnamefont {J.~D.}\ \bibnamefont
  {Hannukainen}}, \bibinfo {author} {\bibfnamefont {Q.}~\bibnamefont {Marsal}},
  \bibinfo {author} {\bibfnamefont {D.~M.}\ \bibnamefont {noz Segovia}},\ and\
  \bibinfo {author} {\bibfnamefont {A.~G.}\ \bibnamefont {Grushin}},\
  }\bibfield  {title} {\bibinfo {title} {Amorphous topological matter: {T}heory
  and experiment},\ }\href {https://doi.org/10.1209/0295-5075/acc2e2}
  {\bibfield  {journal} {\bibinfo  {journal} {Europhysics Letters}\ }\textbf
  {\bibinfo {volume} {142}},\ \bibinfo {pages} {16001} (\bibinfo {year}
  {2023}{\natexlab{a}})}\BibitemShut {NoStop}%
\bibitem [{\citenamefont {Corbae}\ \emph
  {et~al.}(2023{\natexlab{b}})\citenamefont {Corbae}, \citenamefont {Ciocys},
  \citenamefont {Varjas}, \citenamefont {Kennedy}, \citenamefont {Zeltmann},
  \citenamefont {Molina-Ruiz}, \citenamefont {Griffin}, \citenamefont
  {Jozwiak}, \citenamefont {Chen}, \citenamefont {Wang}, \citenamefont {Minor},
  \citenamefont {Scott}, \citenamefont {Grushin}, \citenamefont {Lanzara},\
  and\ \citenamefont {Hellman}}]{corbae_natmat_2023}%
  \BibitemOpen
  \bibfield  {author} {\bibinfo {author} {\bibfnamefont {P.}~\bibnamefont
  {Corbae}}, \bibinfo {author} {\bibfnamefont {S.}~\bibnamefont {Ciocys}},
  \bibinfo {author} {\bibfnamefont {D.}~\bibnamefont {Varjas}}, \bibinfo
  {author} {\bibfnamefont {E.}~\bibnamefont {Kennedy}}, \bibinfo {author}
  {\bibfnamefont {S.}~\bibnamefont {Zeltmann}}, \bibinfo {author}
  {\bibfnamefont {M.}~\bibnamefont {Molina-Ruiz}}, \bibinfo {author}
  {\bibfnamefont {S.~M.}\ \bibnamefont {Griffin}}, \bibinfo {author}
  {\bibfnamefont {C.}~\bibnamefont {Jozwiak}}, \bibinfo {author} {\bibfnamefont
  {Z.}~\bibnamefont {Chen}}, \bibinfo {author} {\bibfnamefont {L.-W.}\
  \bibnamefont {Wang}}, \bibinfo {author} {\bibfnamefont {A.~M.}\ \bibnamefont
  {Minor}}, \bibinfo {author} {\bibfnamefont {M.}~\bibnamefont {Scott}},
  \bibinfo {author} {\bibfnamefont {A.~G.}\ \bibnamefont {Grushin}}, \bibinfo
  {author} {\bibfnamefont {A.}~\bibnamefont {Lanzara}},\ and\ \bibinfo {author}
  {\bibfnamefont {F.}~\bibnamefont {Hellman}},\ }\bibfield  {title} {\bibinfo
  {title} {Observation of spin-momentum locked surface states in amorphous
  {Bi$_2$Se$_3$}},\ }\href {https://doi.org/10.1038/s41563-022-01458-0}
  {\bibfield  {journal} {\bibinfo  {journal} {Nature Materials}\ }\textbf
  {\bibinfo {volume} {22}},\ \bibinfo {pages} {200} (\bibinfo {year}
  {2023}{\natexlab{b}})}\BibitemShut {NoStop}%
\bibitem [{\citenamefont {Costa}\ \emph {et~al.}(2019)\citenamefont {Costa},
  \citenamefont {Schleder}, \citenamefont {Buongiorno~Nardelli}, \citenamefont
  {Lewenkopf},\ and\ \citenamefont {Fazzio}}]{fazzio_nanolett_2019}%
  \BibitemOpen
  \bibfield  {author} {\bibinfo {author} {\bibfnamefont {M.}~\bibnamefont
  {Costa}}, \bibinfo {author} {\bibfnamefont {G.~R.}\ \bibnamefont {Schleder}},
  \bibinfo {author} {\bibfnamefont {M.}~\bibnamefont {Buongiorno~Nardelli}},
  \bibinfo {author} {\bibfnamefont {C.}~\bibnamefont {Lewenkopf}},\ and\
  \bibinfo {author} {\bibfnamefont {A.}~\bibnamefont {Fazzio}},\ }\bibfield
  {title} {\bibinfo {title} {{Toward Realistic Amorphous Topological
  Insulators}},\ }\href {https://doi.org/10.1021/acs.nanolett.9b03881}
  {\bibfield  {journal} {\bibinfo  {journal} {Nano Letters}\ }\textbf {\bibinfo
  {volume} {19}},\ \bibinfo {pages} {8941} (\bibinfo {year}
  {2019})}\BibitemShut {NoStop}%
\bibitem [{\citenamefont {Huang}\ and\ \citenamefont
  {Liu}(2018{\natexlab{b}})}]{huang_prl_2018}%
  \BibitemOpen
  \bibfield  {author} {\bibinfo {author} {\bibfnamefont {H.}~\bibnamefont
  {Huang}}\ and\ \bibinfo {author} {\bibfnamefont {F.}~\bibnamefont {Liu}},\
  }\bibfield  {title} {\bibinfo {title} {Quantum {S}pin {H}all {E}ffect and
  {S}pin {B}ott {I}ndex in a {Q}uasicrystal {L}attice},\ }\href
  {https://doi.org/10.1103/PhysRevLett.121.126401} {\bibfield  {journal}
  {\bibinfo  {journal} {Phys. Rev. Lett.}\ }\textbf {\bibinfo {volume} {121}},\
  \bibinfo {pages} {126401} (\bibinfo {year} {2018}{\natexlab{b}})}\BibitemShut
  {NoStop}%
\bibitem [{\citenamefont {Prodan}\ and\ \citenamefont
  {Prodan}(2009)}]{prodan_phonons_prl2009}%
  \BibitemOpen
  \bibfield  {author} {\bibinfo {author} {\bibfnamefont {E.}~\bibnamefont
  {Prodan}}\ and\ \bibinfo {author} {\bibfnamefont {C.}~\bibnamefont
  {Prodan}},\ }\bibfield  {title} {\bibinfo {title} {{Topological Phonon Modes
  and Their Role in Dynamic Instability of Microtubules}},\ }\href
  {https://doi.org/10.1103/PhysRevLett.103.248101} {\bibfield  {journal}
  {\bibinfo  {journal} {Phys. Rev. Lett.}\ }\textbf {\bibinfo {volume} {103}},\
  \bibinfo {pages} {248101} (\bibinfo {year} {2009})}\BibitemShut {NoStop}%
\bibitem [{\citenamefont {Süsstrunk}\ and\ \citenamefont
  {Huber}(2015)}]{mechanical_topology_science}%
  \BibitemOpen
  \bibfield  {author} {\bibinfo {author} {\bibfnamefont {R.}~\bibnamefont
  {Süsstrunk}}\ and\ \bibinfo {author} {\bibfnamefont {S.~D.}\ \bibnamefont
  {Huber}},\ }\bibfield  {title} {\bibinfo {title} {Observation of phononic
  helical edge states in a mechanical topological insulator},\ }\href
  {https://doi.org/10.1126/science.aab0239} {\bibfield  {journal} {\bibinfo
  {journal} {Science}\ }\textbf {\bibinfo {volume} {349}},\ \bibinfo {pages}
  {47} (\bibinfo {year} {2015})}\BibitemShut {NoStop}%
\end{thebibliography}%

\end{document}